\documentclass[apj]{emulateapj}
\usepackage{lscape}

\def\lsim{\lower.5ex\hbox{$\; \buildrel < \over \sim \;$}}
\def\gsim{\lower.5ex\hbox{$\; \buildrel > \over \sim \;$}}
\def\abeq{\lower.7ex\hbox{$\; \buildrel \sim \over - \;$}}

\def\t{\ifmmode {\tau} \else $\tau$ \fi}

\def\ref{\noindent \hangafter=1 \hangindent=0.7 truecm}

\def\cm{\ifmmode {\rm cm}^{-1} \else cm$^{-1}$ \fi}
\def\s{\ifmmode {\rm s}^{-1} \else s$^{-1}$ \fi}
\def\cc{\ifmmode {\rm cm}^{-3} \else cm$^{-3}$ \fi}
\def\cs{\ifmmode {\rm cm}^{-2} \else cm$^{-2}$ \fi}
\def\g{\ifmmode \gamma \else $\gamma$\fi}
\def\G{\ifmmode \Gamma \else $\Gamma$\fi}

\def\kms{\ifmmode {\rm km\ s}^{-1} \else km s$^{-1}$\fi}

\usepackage{graphicx}

\begin{document}

\title{The X-ray Emissions from the M87 Jet:  Diagnostics and Physical 
Interpretation}

\author{Eric S. Perlman\altaffilmark{1,2}}

\altaffiltext{1}{Department of Physics, Joint Center for Astrophysics,
University of Maryland-Baltimore County, 1000 Hilltop Circle, Baltimore, MD
21250, USA}

\altaffiltext{2}{Department of Physics and Astronomy, Johns Hopkins University,
3400 North Charles Street, Baltimore, MD 21218, USA}

\author{Andrew S. Wilson\altaffilmark{3,4}}
 
\altaffiltext{3}{Astronomy Department, University of Maryland, College Park, 
MD 20742, USA}

\altaffiltext{4}{Space Telescope Science Institute, 3700 San Martin Drive, 
Baltimore,MD  21218, USA}

\email{perlman@jca.umbc.edu, wilson@astro.umd.edu}

\begin{abstract}

We reanalyze the deep {\it Chandra} observations of the M87 jet, first 
examined by Wilson \& Yang (2002).  By employing an analysis chain that also
includes image deconvolution, knots HST-1 and I are fully separated from
adjacent emission.  We derive the spatially resolved X-ray spectrum of the jet
using the most recent response functions, and find slight  but significant
variations in the spectral shape, with values of $\alpha_x (S_\nu \propto
\nu^{-\alpha})$ ranging from $\sim 1.2-1.4$ (in the nucleus, knots HST-1, D and
C) to $\sim 1.6$  (in knots F, A and B).  We make use of VLA radio
observations, as well as {\it HST} imaging and polarimetry data (Perlman et al.
1999, 2001a), to examine the jet's broad-band spectrum and inquire as to the
nature of particle acceleration in the jet.
As shown in previous papers,
a simple continuous injection
model for the synchrotron-emitting knots, in which one holds constant both the
filling factor, $f_{acc}$, of the regions within which particles are
accelerated and the energy spectrum of the injected particles, cannot account
for the flux or spectrum of the X-ray emission. 
Instead, we propose that $f_{acc}$ is a
function of both position and energy  and find that in  the inner jet, $f_{acc}
\propto E_\gamma^{-0.4 \pm  0.2} \propto E_e^{-0.2 \pm 0.1}$, and in knots A
and B, $f_{acc} \propto E_\gamma^{-0.7 \pm  0.2} \propto E_e^{-0.35 \pm 0.1}$,
where $E_\gamma$ is the energy of the emitted photon and $E_e$ is the energy of
the emitting electron.
In this model, the index, $p$, of the relativistic electron energy spectrum
at injection ($n(E_{e}) \propto E_{e}^{-p}$) is $p=2.2$ at all energies and
all locations along the jet, in excellent agreement with the predictions of
models of cosmic ray acceleration by ultrarelativistic shocks ($p$=2.23).
There is a strong correlation between the peaks of
X-ray emission and minima of optical percentage polarization, i.e., regions
where the jet magnetic field is not  ordered.  We suggest that the X-ray peaks
coincide with shock waves which  accelerate the X-ray emitting electrons and
cause changes in the direction of the magnetic field; the polarization is thus
small because of beam averaging.

\end{abstract}


\section{Introduction}

The giant elliptical galaxy M87 hosts the best-known extragalactic jet.  As a
result of its proximity (distance = 16 Mpc, Tonry 1991, for a scale of $1'' =
78$ pc) and high surface brightness from radio through X-rays, particularly
high resolution studies of its structure are possible.  The synchrotron nature
of the jet's optical emissions was first demonstrated by Baade (1956), who
found it to be highly polarized. The first high-quality imaging polarimetry and
photometry was done in the 1970s, both in the radio (De Young, Hogg \& Wilkes
1979) and optical (Schmidt, Peterson \& Beaver 1978).  Those observations
showed a magnetic field predominantly parallel to the local jet direction,
highlighting the importance of an understanding of the field  structure. 
Subsequent observations in the radio (Owen, Hardee \& Cornwell 1989 and
references therein; Biretta et al. 1995; Zhou 1998), mid-IR (Perlman et al.
2001b, hereafter P01b), and near-IR/optical (Meisenheimer, R\"oser \&
Schl\"otelburg 1996; Sparks, Biretta \& Macchetto 1996; Perlman et al. 1999,
hereafter P99; Perlman et al. 2001a, hereafter P01a) have featured detailed
studies of the jet's structure, allowing models of synchrotron emission to be
fit to the broadband spectrum of each knot in the jet, and also allowing the
axial and magnetic field structure to be probed.

Given its high surface brightness, it is not surprising that the jet of M87 was
also one of the first three jets detected in the X-ray band by {\it Einstein}
(Schreier et al. 1982; Biretta, Stern \& Harris 1991), along with those of Cen
A (Schreier et al. 1979; Feigelson et al. 1981) and 3C 273 (Harris \& Stern
1987).  However, prior to the launch of {\it Chandra} in 1999, very little was
known about its X-ray structure.  X-ray emission had been discovered from three
of the knots in the jet by Biretta et al. (1991), but due to the low angular
resolution of the {\it Einstein} and {\it ROSAT} data ($\gsim 5''$, compared to
jet features which are $\sim 1''$ in extent), it was not possible to make
detailed
analyses of the jet's X-ray morphology, let alone comparisons to the radio or
optical.

{\it Chandra} observations of the M87 jet have revealed X-ray emissions from
every knot in the M87 jet (Marshall et al. 2002, hereafter M02; Wilson \& Yang
2002, hereafter WY02).   The X-ray spectra of the knots are steeper than those
in the radio and optical bands, confirming that the X-ray emission is not
inverse-Compton scattering, for which the same spectral index is expected for
the synchrotron and inverse-Compton emission from the same electron
population.   M02 and WY02 argued that the X-ray emission represents the
high-energy tail of the synchrotron spectrum. However, the images show
intriguing differences between the X-ray morphology of the M87 jet and its
morphology in the optical and radio: in particular, the X-ray emissions from
some knots appear to be upstream of their optical emissions and, in addition,
two X-ray bright regions
(labelled 'D-X' and 'G' by M02)
are seen, which do not coincide with optically
bright regions of the jet.
By contrast, the radio and optical morphologies of
the M87 jet differ only subtly, being quite similar on arcsecond scales but
somewhat narrower and ``knottier'' in the optical as revealed by the HST
observations (Sparks et al. 1996).   

X-ray synchrotron emitting electrons have $\gamma = E/mc^2 \sim 10^{7-8}$ and
radiative lifetimes of a few to tens of years (given the magnetic field
strengths estimated by Meisenheimer et al. 1996 or Heinz \& Begelman 1997),
thus requiring local particle acceleration.  Therefore, {\it Chandra} images of
M87 give us the opportunity to probe directly the acceleration of the
high-energy particles whose emissions we see. In light of this fact, we decided
to analyze together the deepest published {\it Chandra} image of the M87 jet
(WY02) and the multiwaveband VLA and HST imaging and polarimetry published by
Perlman et al. (P99, P01a).  The goal of this work is to improve our
understanding of the physics of the M87 jet, and the relationship between the
X-ray and optical emissions.  

In \S 2 we review the observations used in each band.  In \S 3 we discuss the
radio, optical and X-ray morphology and spectrum of the jet.  This discussion
includes a presentation of deconvolved images made from the {\it Chandra} X-ray
Observatory data and a comparison with HST images. The deconvolution results
in a significant gain in resolution which proves quite helpful in understanding
several structural details.  We also present in \S 3 a detailed discussion of
the jet's X-ray and optical-to-X-ray spectrum, including the first
optical-to-X-ray spectral index map as well as a comparison
between the jet's X-ray structure and  polarized optical emissions. In \S 4 we
discuss the physical implications of our results, including a model for
particle acceleration and insights into the  jet structure and magnetic field
configuration in the M87 jet.  We conclude our discussion in \S 5 with a
summary.

\section {Observations}

The data used in this paper have already been discussed by previous
authors: the {\it Chandra} data by WY02 and the VLA and {\it HST} data by P99
and P01a (see also Zhou 1998 for more details on the VLA observations).  In
this section, we review the essentials of each dataset.  For specific
information on data reduction procedures (except those noted below, which are
specific to this paper) or observational details we refer the reader to those
papers.

\subsection{{\it Chandra} Observations}

The {\it Chandra} data cover the energy range $0.2-10$ keV, corresponding to
frequencies between $4.8 \times 10^{16}$ and $2.4 \times 10^{18}$ Hz.  The {\it
Chandra} observations were taken on 2000 July $29-30$, using the ACIS-S
instrument, with the entire jet on chip S3.  The observations were split into
two parts after a preliminary 1 ks observation showed that at least three
components in M87's X-ray structure would be piled up in a normal 3.2s frame 
time observation.  A 0.4s frame-time observation with a standard 1/8 subarray
was used to obtain accurate spectra and photometry of the brightest components,
while a 3.2s frame-time  observation using the full S3 CCD (as well as CCDs I2,
I3, S1, S2 and S4) allowed fainter structure to be observed (WY02).  The good
exposure time of the 0.4s frame-time observation was 12.8 ks, while the 3.2s
frame-time data included 37.6 ks of good exposure time. 

For data reduction, we followed the procedures recommended in the CIAO science
threads.   The data were filtered to exclude times of high background and
aspect errors, of which very few were found.  Prior to analyzing the {\it
Chandra} data, we resampled the images by a factor of 4, obtaining a scale of
$0\farcs123$/pix.  We also applied pixel derandomization to the imaging data. 
Upon examination of the 3.2s frame time data, four components were found to be
piled up: the nucleus as well as knots HST-1, D and A in the jet (WY02).  For
these components, it was therefore necessary to use the 0.4s frame-time
observations for all analysis.  For all other regions of the jet, however, we
were able to make use of the higher sensitivity 3.2s frame time data.  All data
selection for X-ray spectral analysis was done in CIAO v2.2 and 2.3, following
the appropriate threads.  We included the application of ACISABS to the ARF,
which corrects for absorption by contaminants on the surface of the CCD or a
pre-CCD filter.   For all jet regions, we used X-ray background data selected
from $25''\times 3''$ rectangular regions situated $5''$ north and south of the
jet.  

For the purposes of analyzing the imaging data, we made use of a hybrid
strategy, screening the images to select only the 0.4s frame-time data for
pixels within $1\farcs5$ of the centroids of the knots which are piled up
in the 3.2s frame-time data, while using a
weighted average of the 3.2s and 0.4s frame-time datasets in all other regions.
In this way we were able to make an X-ray map of the jet which is essentially
unaffected by pileup.   A short IDL program was used to do the data selection
described above. Using this data selection method, maps were made in the
$0.3-1.0$ keV, $1-3$ keV and $3-10$ keV bands as well as the $0.3-1.5$ keV (see
below) and total bands.  These maps were then registered to the VLA data by
assuming that the nuclear emission peak is located at the same place in both
bands.

To analyze the jet's subarcsecond morphology, we attempted to deconvolve the
PSF from the {\it Chandra} image.  This is a difficult task, since the PSF of
{\it Chandra} is known to be energy-dependent (e.g., Karovska  et al. 2001). 
Fortunately, the energy-dependent variations in the PSF size can be minimized
by choosing for deconvolution only X-ray photons with energies between 0.3 and
1.5 keV.  At higher energies the PSF has stronger wings due to scattering by
the mirror, and the intensity of these wings is energy dependent. We therefore
obtained a $0.3-1.5$ keV image, using the data selection described above, and
attempted deconvolution using a monochromatic, 1 keV PSF, following the recipe
in the ``Create a PSF'' CIAO 2.2 thread.  Several deconvolution algorithms were
tried, both in AIPS and IRAF.  The most satisfactory results were obtained with
the maximum entropy algorithm VTESS in AIPS.  Other algorithms, including
the tasks LUCY in IRAF and IMAGR in AIPS, gave similar results in terms of the
source structure but had less even and/or higher noise levels across the
image.  To reach a stable solution required 150 iterations  of VTESS.
Following the application of VTESS, we smoothed the data with a Gaussian of
FWHM  $0\farcs2$.

\subsection{HST Observations}

The {\it HST} images we discuss in this paper include both multi-band imaging
and polarimetry.  The {\it HST} data cover the frequency range $1.5 - 10 \times
10^{14} $ Hz.

The multi-band {\it HST} imaging data (P01a) were obtained in 1998 with the
WFPC2 and NICMOS instruments.  Data were obtained on 25 and 26 February through
six filters, spanning the wavelength range 0.3-1.6 microns.  A seventh near-IR
band (2.05 microns) was also scheduled for 26 February, but had to be
reobserved on 4 April 1998, due to a loss of guide-star lock.  Because of the
relatively small field of view of the {\it HST} instruments, we performed the
1998 February observations with the {\it HST} oriented such that the jet fell
along a chip diagonal.  Unfortunately an equally good orientation was not
available on 4 April; as a result, the F205W data do not include the region
within $1''$ of the nucleus.  In the F110W and F160W bands, the small size of
the NIC1 detector made it necessary to observe at two positions.  Most of the
observing time for those bands was used on the inner jet (i.e., features
interior to knot A, at a distance from the nucleus of $r=12\farcs4$) because
it has a lower optical surface brightness than the main features of the outer
jet (i.e., beyond knot A).

Optical polarimetry of the M87 jet (P99) was obtained on May 27, 1995 with the
{\it HST}, using WFPC2 and the F555W (broadband V) filter plus the POLQ
polarization quad filter.  In order to maximize the unvignetted field of view
available for these observations, we used the WF chips, rather than the PC, for
the polarization observations.  This is important for the M87 jet, which
is over $20''$ long: by comparison the unvignetted field of view on the PC chip
using the POLQ filter is only $15''$ square and while most, if not all, of the
jet could be included on the unvignetted part of the PC, it would be difficult
if not impossible to obtain in this way the good subtraction of the galaxy
emission which is critical for polarimetry.  Using each of the WF chips with
POLQ is equivalent to obtaining images at $PA = 45^\circ - V3, 90^\circ - V3$,
and $135^\circ - V3$, where $V3$ is the rotation of the HST's z-axis with
respect to the sky (for details, see Biretta \& McMaster 1997).  A total of
1800 seconds integration time, split in three to reject cosmic rays, was
obtained on each WF chip.

Each {\it HST} dataset was reduced in IRAF using the best recommended
flatfields, biases, darks and illumination correction images. Combination of
the {\it HST}/WFPC2 data was done in CRREJECT to eliminate cosmic rays.  The
F814W images, as well as all NICMOS images, were combined using the tasks in
the DRIZZLE package. The best available geometrical correction files
were used to make minor geometric corrections essential to registering the HST
data to the radio and X-ray frames of reference.  Prior
to analyzing the emissions of the jet, we subtracted the galaxy emission, using
the tasks ELLIPSE, BMODEL and IMCALC.  All flux calibration was done using
SYNPHOT in IRAF.   As  described in P99 and P01a, we registered all {\it HST}
data to the VLA data by assuming that the radio and optical core positions are
identical (i.e., the same method as used for the {\it Chandra}
data).  In comparing the morphology and spectral properties of the jet in the
optical to those in the X-rays, we smoothed the data with Gaussians to
$0\farcs5$ resolution to ensure that no effects were present from 
observing with 
different resolutions.    The polarimetry data were, however, left at the
original resolution of $0\farcs2$ to give additional insights into the
physics.  For more details, we refer the reader to P99 and P01a.

\subsection{VLA Data}

To analyze the jet's broadband spectrum, we include also previously published
radio data (Zhou 1998, P99), obtained in February 1994 at 15 GHz.  The data
reduction procedures for this dataset have been described in Zhou (1998) and
P99.  To facilitate direct comparison between the radio and optical images, the
radio image was resampled to the same pixel scale as the {\it HST} and {\it
Chandra} images.  As the absolute astrometry from VLA data is of
much higher quality than that which can be attained from either {\it HST} or
{\it Chandra} data, the position of the nucleus of M87 in the VLA data was 
used as the absolute astrometric standard (see previous sections).

While the radio data were obtained in February 1994, the likely effects of
variability are small.  Even for the largest radio variability found between
1993 and 1997 ($\sim 40$\%  for HST-1; Zhou 1998), the effect on $\alpha_{ro}$
is only $\Delta \alpha_{ro}=0.04$. As with the HST data, we have convolved the
VLA data with a Gaussian to obtain $0\farcs5$ resolution.

\section{The Morphology and Spectrum of the M87 jet}

Here we compare the best available images and spectra of the jet at radio,
optical and X-ray frequencies.  Below, we first discuss the jet's X-ray
morphology and its relationship to the optical and radio morphology (\S 3.1),
then its optical-to-X-ray spectral morphology (\S 3.2), and then a comparison
of its optical polarimetry and X-ray morphology (\S 3.3). 

\subsection{Radio, Optical and X-ray Morphology in Total Flux}

In Figure 1 we compare radio, optical and X-ray images of the M87 jet with
images of the radio-optical, optical and optical-to-X-ray spectral indices
($\alpha_{ro}$, $\alpha_o$ and $\alpha_{ox}$ respectively).   For the purpose
of display, all  images in Figure 1 were rotated so that the jet is along the
X-axis. Also noted on Figure 1 are the historical names for regions in the M87
jet, as in P01a and earlier papers.  The X-ray image shown in Figure 1 is that
produced using the deconvolution procedure described in \S 2.1.   As already
noted in WY02, the {\it Chandra} image shows X-ray emission from every knot in
the jet as well as inter-knot regions.  We attempted to improve the signal to
noise in  faint regions by applying adaptive smoothing; however, the
improvement was modest at best so we do not show that version of the image. 

The sub-arcsecond resolution of {\it Chandra} is readily apparent in Figure 1;
the nucleus has a full-width at half-maximum of $0\farcs54$ (perpendicular to
the jet direction) in the deconvolved image, compared to $0\farcs84$ before
deconvolution.  Thus the maximum entropy deconvolution we applied to these data
improved the angular resolution by $\approx 35$\%.  This has an immediate
effect on our ability to separate important regions of the jet.  For example,
unlike  the undeconvolved images presented in WY02 and M02, here we fully
separate the nucleus from HST-1, an X-ray and optically bright knot located
only $0\farcs9$ from the nucleus.  Similarly, we separate more clearly knots I
and A, which are $1''$ apart and  were not well resolved in undeconvolved
images.  

 The improved resolution also allows us to make out several features
which were not immediately apparent in the figures presented by WY02 and M02. 
Knot D is seen to have significant X-ray structure downstream of its peak; in
particular, there does appear to be an X-ray flux enhancement  corresponding to
the radio/optical feature known as D-W.  Figure 1 also shows that there is
indeed a weak optical peak associated with the feature called D-X by M02, which
was also noted but not named by WY02.  Comparison with the optical image
reveals that this feature is most likely associated with the upstream end of
the knot E region, as its optical and radio counterpart is connected to knot E's
optical and radio maxima by an apparent thin bridge.  Another feature is
apparent further downstream, in knot B1, where the X-ray bright region appears
to be associated with only one of the two apparently crossing filaments seen in
the optical and radio (compare with Figure 1 of P99).

In Figure 2, we plot the profiles of various jet components in the direction
perpendicular to the jet, in the radio, optical and X-rays.  In each panel,  we
also plot, for comparison, the profile of the {\it Chandra} PSF, represented by
that of the nucleus, which is assumed to be unresolved.  The profiles in Figure
2 were made by extracting slices 7 pixels (0\farcs86) wide at the centroid of
the named jet component (see Table 1). Three things are apparent from Figure
2.  First, it is clear that we do resolve in the transverse direction several
regions in the jet with {\it Chandra}.  In particular, the transverse X-ray
profiles of knots A, B and C  are wider than the PSF at high significance. 
There are also indications that the X-ray profiles of knot I, E and possibly
knots D-X, D-W and D-E may be resolved in the transverse direction.  The latter
results should be verified with deeper imaging data.   Second, in the regions
where we clearly resolve the X-ray emission, the X-ray flux profile is narrower
than that measured in the optical and radio bands.  Third, the loci of the flux
maxima of each knot are not identical -- i.e., there are real offsets between
the X-ray and radio/optical component maxima in both the  direction parallel to
the jet (as found by M02 and WY02, see also below), and perpendicular to it.

Given that {\it Chandra}'s angular resolution is just sufficient to resolve the
jet transverse to its axis downstream of knot A, it is best to compare the
jet's radio, optical and X-ray morphology in the context of runs of flux along
the jet.  We show these in Figure 3.  As can be seen by examining these plots,
the radio, optical and X-ray flux track one another fairly closely in most
regions of the jet.   However, there do appear to be some differences in
morphology.   We have attempted to quantify these differences by fitting
Gaussians at the location of each flux maximum to the (convolved) optical and
(deconvolved) {\it Chandra} datasets, using the task JMFIT in AIPS.  The
results are given in Table 1.  
%Note that Table  1 gives not only RA and DEC
%position for each component's flux maximum, but also the size of the fitted
%Gaussian and its size when deconvolved.   Table 1 thus gives us important data
%as to the size of each X-ray emission region.

The columns in Table 1 give, respectively, (1) Component Name; (2) RA, as
derived from the {\it Chandra} X-ray data, after assuming the VLA position for
the nucleus; (3) Declination, as derived from the {\it Chandra} X-ray data,
after assuming the VLA position for the nucleus; (4) observed (i.e., not 
deconvolved from the PSF) component major
axis (arcseconds), as derived from the {\it Chandra} X-ray data; (5) observed
component minor axis (arcseconds), as derived from the {\it Chandra} X-ray
data; (6) Distance of component from the nucleus, in arcseconds, as derived
from the {\it Chandra} X-ray data; (7) RA, as derived from the {\it HST} image,
after assuming the VLA position for the nucleus; (8) Declination, as derived
from the {\it HST} data, after assuming the VLA position for the nucleus; (9)
The difference in RA, $\Delta$RA, in arcseconds, in the sense optical$-$X-ray; 
thus a negative value in this column means that the X-ray component is closer
to the nucleus than the optical;
(10) The difference in Declination, $\Delta$Dec, in arcseconds, in  the sense
optical$-$X-ray; (11) Any notes connected with the astrometry (legend given at
bottom).  The numbers given in parentheses below each parameter represent 
the error in that parameter.  

The data in Table 1 confirm that we resolve two jet components -- namely
knots A and B.  However, they neither support nor reject the claims we make for
knots C and I, as the surface brightnesses of those features are small enough
that it was necessary to fix the size of the Gaussians used to fit them to
obtain a good fit.  Two other regions appear to have large Gaussian sizes as
results of this procedure, namely knots E and F.  However we cannot claim to
have resolved these clearly, as we did not restrict the size of the Gaussians
in the direction perpendicular to the jet,  and moreover as seen in Figure 2,
their profiles are not conclusive as to whether the {\it Chandra} data resolve
them.  These same tests should be redone on knots E and F with summed {\it
Chandra} data.

As can be seen in Table 1, most components do not have significantly different
positions for their X-ray and optical flux maxima.   There are, however,
significant ($>3 \sigma$ and $>0\farcs1$) optical-X-ray offsets for  4
components. These are knots  D-W, D-X, F and C-2. Two other knots, HST-1 and A,
have offsets that are $>3 \sigma$ but $\approx 0\farcs1$ .  Because of the
small sizes of these offsets (i.e., $<0.2 \times$ the angular resolution  of
the {\it Chandra} image), these offsets should be treated as less secure than
larger offsets with identical statistical significance. The other knots have
offsets in the range 1-3 $\sigma$ which are not statistically significant.  We
deal with each of these in turn.

Knot HST-1 is the region of the jet where the fastest apparent superluminal
motions have been seen ($\approx 6 c$, Biretta et al. 1999).  It is also the
location of a bright  flare during 2002-2003 (Harris et al. 2003, Perlman et
al. 2003), at which times the optical and X-ray flux maximum positions did
coincide.  M02 were also not able to find any optical-X-ray offsets; however,
they did not apply deconvolution or pixel derandomization and therefore their
reduction was not as sensitive to small offsets.  Given the X-ray and optical
flaring activity that is currently occurring in this region (Harris et al.
2003, Perlman et al. 2003, Biretta et al. 2004), and which began in $\sim 1999$
(i.e.,  between the {\it HST} and {\it Chandra} observations),  we speculate
that this possible optical-X-ray offset is a result of the flare.

Knots D-W and D-X are rather different cases.   Knot D-W is not associated
with a clear X-ray maximum, whereas there is a clearer optical maximum, as can
be seen in both Figures 1 and 3.   Knot D-X is in just the opposite  situation
-- while there is a clear X-ray  maximum at that location there is not a clear
optical maximum.   In both cases it is therefore not surprising that we see
some X-ray-optical offset.  Our knowledge of the X-ray properties of these
components would clearly benefit from deeper {\it Chandra} observations.

Knot F has two fairly clear optical maxima, which are separated by about
$0\farcs4$ in the HST images (see e.g., P01a, Sparks et al. 1996); the
optically brighter component is the one further downstream.   When the optical
data are convolved to the resolution of {\it Chandra}, one cannot separate by 
eye these components.  It is  apparent that the X-ray maximum of knot F is
associated with the upstream optical maximum.  This observation agrees with
WY02 (M02 did not have good enough statistics to comment on this region, as
they discuss).

Our data suggest a small ($\sim 0\farcs1$) but statistically significant  
($3.5 \sigma$ in RA; $5.5 \sigma$ in Dec) offset between the location of the
optical and X-ray maxima of knot A.  In this we agree with M02; we are also not
in disagreement, however, with WY02, who note no offset to within $\pm
0\farcs1$ but did not analyze HST data.  The origin of this possible offset is
not clear; however, it may have to do with the knot's double structure as seen
at $0\farcs1$ resolution in the optical (P01a, Sparks et al. 1996).  Indeed, a
closer inspection of Figure 1 shows that the flux maximum region of knot A may 
be extended in the X-rays, along the length of the jet.  This region would
clearly benefit from improved angular resolution in the X-rays.

In the knot C region there also appear to be significant X-ray -- optical 
differences, but the situation is somewhat complex.  The X-ray and optical
maxima of the region known in the optical as C-1 appear to coincide.  However,
knot C is a rather diffuse region in the optical, while in the X-rays there are
two distinct maxima (Fig. 3). The second of these is close to the position of
the optical component called  C-2, but it is offset from the maximum of that
component by about $0\farcs7$.   This was noticed also by M02 (who called the
X-ray emission 'G'; however the optical component by that name is located
further from the nucleus) and WY02; however, neither of those teams gave a 
value for this offset.  The $\alpha_{ox}$ map also shows that the downstream
edge of knot C appears to have  a smaller  $\alpha_{ox}$ than regions further
upstream. This region would benefit from analysis of deeper Chandra
observations.

Two other regions are also of note here.  As can be seen in Table 1, our data
suggest optical-X-ray offsets of nearly $3 \sigma$  significance at the
flux maxima of knots E and D-E.  The former was noted by WY02 but not M02; our
data are suggestive of the same optical-X-ray offset (in the sense that the
X-ray maximum is slightly upstream of its optical counterpart).  These regions
would clearly benefit from greater resolution in the X-rays given that the
possible offsets are $\sim 0\farcs1$, only 1/5 of the {\it Chandra} pixel size.

\subsection{Optical-to-X-ray, X-ray and Broadband Spectrum of the M87 Jet}

These data allow us to deduce new information regarding the X-ray and broadband
spectrum of the M87 jet.  M02 and WY02 published the first X-ray spectra of
individual jet components.  Those works established that the X-ray spectrum of
each knot in the jet could be well described by  a steep power law 
($\alpha_x>1$), although no information was included on knot HST-1, in either
paper.  

We have re-analyzed the X-ray spectrum of various components of the M87 jet,
using regions similar to those in WY02, with the addition of a region for knot
HST-1.  Our analysis procedure included two  differences from WY02.  We made
use of the latest available reduction procedures (see \S 2.1), including the
application of ACISABS.   To extract the background spectrum, we used
rectangles parallel to (north and south of) the jet to extract the background
spectrum, a slightly different strategy from that used by WY02 or M02.  We
fitted single power law plus Galactic absorption models for all components, and
in separate trials allowed first only $\alpha_x$ and then both $\alpha_x$ and
$N_H$ to vary.  The full analysis was also done separately by each author, with
ESP using Sherpa and ASW using XSPEC for spectral modelling. Data were
generally fit in the $0.3-5$ keV band. The results of these procedures are
given in Table 2.

All knots are well described by single power law models; none requires a
significant spectral break within the X-ray band.  The variations in the
power-law index of the jet's X-ray spectrum are small; in fact, all the knots
appear consistent to within $2 \sigma$ with a  power law of energy index
$\alpha \sim 1.45$.  There is, however, evidence for X-ray spectral variations
along the jet.  We find knot HST-1  to have an X-ray spectral index $\alpha_x
\approx 1.3$, in agreement with WY02's estimate without deconvolution. 

We reproduce to within the 90\% confidence errors most of the values for
$\alpha_x$ given in WY02 and M02 for which fits are given in those papers.  We
also find that the absorbing column densities are consistent with the Galactic
value, $N_H({\rm Gal})=2.4 \times 10^{20} {\rm ~cm^{-2}}$, with weak evidence
that the column to the nucleus may exceed $N_H$(Gal).  There are some areas of
disagreement between our results and those of WY02, however.  In particular, we
do not unambiguously  verify an absorbing column in excess of Galactic for the
nucleus, as found by WY02, although both our analyses give a higher absorbing
column for the nucleus than anywhere else. We also find that the 
photon indices of the knots given by WY02 
are too small by, on average, 0.15--0.2, as noted
in the erratum to WY02 (Wilson \& Yang 2004). Our testing reveals that the
likely cause for these discrepancies is  our correction of the ARF for the
effects of absorption by contaminants on the surface of the CCD or its filter,
using the program ACISABS.  The existence of such absorption was not known when
WY02 analyzed their data, with the results that their spectra for knots D, A
and B are too hard (by $\Delta \alpha \sim 0.2-0.3$) and their column density
of the nucleus is overestimated.  Thus the discussion by WY02 of the putative
hard spectra of D, A and B (their \S 4.2) is not correct.

We also made a profile of $\alpha_x$ along the jet,
by extracting  photons in a $4''$ wide strip and then dividing the regions so
that each had approximately 800 counts. This procedure was done separately for
both the 3.2 s frame time data set and for the 0.4s frame time dataset to allow
for pileup in the bright regions.  The fits were then done with $N_H$ left
constant given the information gleaned from the fits to individual knots. The
run of $\alpha_x$ along the jet derived in this way is shown in the bottom
panel of Figure 4. This map verifies the result above that all jet regions
appear consistent to within $2 \sigma$ with a single power law shape.  We do,
however,  find possible variations at the nearly 2$\sigma$ level, with somewhat
flatter spectra interior to knot E, as well as in knot C, and steeper spectra
in knots F, A and B.  

%Our analysis verifies at 99\% confidence (according to an $F$-test)
%the enhanced absorbing column for the nucleus noted by WY02; the
%column required 

To investigate this issue further, we constructed images of two X-ray softness
ratios $SR1$= F($0.3-1$ keV)/F($1-3$ keV) and $SR2$= F($1-3$ keV)/F($3-10$
keV).  To increase the signal to noise in the $3-10$ keV band (which contains
the smallest number of photons), we performed this calculation on the data
binned to $0\farcs492$ pixels, rather than  $0\farcs123$ pixels, and smoothed
the data with a  Gaussian of $\sigma=1$ pixel.  We plot these ratios in Figure
4 (middle  panels).  As can be seen, the softness ratios are significantly
smaller in the inner $3''$ of the jet, i.e., the nucleus, HST-1 and possibly
D-E, have harder X-ray spectra than regions further out.  We also see evidence
for some flattening of the X-ray spectrum in knot C, particularly in $SR1$.
Thus, the $2 \sigma$ variations found in $\alpha_x$ (shown in the bottom panel
of  Figure 4) are supported by the softness ratio analysis and are likely to be
real.

A map of optical-to-X-ray spectral index, $\alpha_{ox}$, is shown in Figure 1,
bottom panel.  The run of $\alpha_{ox}$ along the jet is shown in Figure 5
compared to runs of $\alpha_o$ and $\alpha_{ro}$.  The $\alpha_{ox}$ map was
made from the deconvolved, pileup-corrected 0.3--1.5 keV {\it Chandra} image and
the F300W {\it HST} image from P01a, convolved to a resolution of $0\farcs5$.
Making the $\alpha_{ox}$ map required converting the {\it Chandra} image into
units of $\mu$Jy per pixel at 1 keV, which was done assuming a constant
spectral index $\alpha_x = 1.4$ (see above) and the Galactic column density,
$N_H({\rm Gal}) = 2.4 \times 10^{20} {\rm~cm^{-2}}$ (Biretta et al. 1991).  The
uncertainty in $\alpha_{ox}$ resulting from the assumption of constant
$\alpha_x$ and $N_H$ is tiny. The runs of $\alpha_o$ and $\alpha_{ro}$  are
taken from P01a and smoothed to $0\farcs5$ resolution.

As can be seen from Figures 4 and 5, $\alpha_{ox}$ varies considerably along 
the M87 jet.  In knot HST-1, $\alpha_{ox}$ is considerably smaller than
in any other region of the jet, and is consistent with its large X-ray to
optical ratio noted by M02 and WY02.  The value of  $\alpha_{ox}$ (HST-1) is
0.83, as compared to 1.45 for the nucleus and 1.2 for knot D.   Note, however,
that  due to variability (see above), this value of $\alpha_{ox}$ may not be
reliable. Beyond knot HST-1, there are two types of variation seen in
$\alpha_{ox}$.  First, there is a steady spectral steepening, which begins at
$2''$ from the nucleus (the inter-knot region between knots HST-1 and D) and
extends to $18''$ from the nucleus (in knot C, between the two X-ray peaks). 
In this region  $\alpha_{ox}$ increases from $\approx 1.2-1.4$ at $2''-7''$
from the nucleus (knots D, E) to $\approx 1.7-1.9$ at $15''-18''$ from the
nucleus (knots B, C).    Second, superposed on this steady increase are small,
but significant, variations in the optical-to-X-ray spectrum at the positions
of optical and X-ray flux maxima:  at the positions of X-ray and optical maxima
in the inner jet,  $\alpha_{ox}$ decreases.    These variations in 
$\alpha_{ox}$ appear largely to mirror the variations in $\nu_{break}$ 
determined by P01a through fitting the radio and optical spectra. 

\subsection{Comparison of X-ray Morphology with Optical Polarimetry}

A comparison between the  morphology of the X-ray emission and maps of optical
polarized flux and polarization position angle,  can give important information
on the configuration of the magnetic field in the X-ray emitting regions.  This
comparison is shown in Figure 6.  The top panel of Figure 6 displays the X-ray
flux (greyscale) with optical percentage polarization contours overplotted.  On
this panel, the datasets  have been rotated so that the jet is along the
X-axis, as in Figures 1-5.  The lower three panels of Figure 6 show X-ray flux
contours plotted along with  vectors whose direction is that of the magnetic
field, as derived from the  optical polarization measurements, and whose length
is proportional to the  optical percentage polarization.   These panels are in
the cardinal orientation (i.e., north up, east to the left).    Each of the
three lower panels shows a different region of the jet, with the region out to
5$''$ from the nucleus (knots HST-1, D-E, D-W and D-X)  shown in the left
middle panel, the region between knot E and knot I ($4\farcs5- 11\farcs5$ from
the nucleus) shown in the right middle panel, and the outer jet (knot A and 
exterior to it, $11\farcs5-19''$ from the nucleus) shown in the lower panel.

Figure 6 reveals a fairly consistent anticorrelation between optical percentage
polarization and X-ray flux in the inner jet.  In particular, the bright X-ray
flux peaks are generally regions of low optical polarization, while peaks of
optical polarization occur well away from the peaks of X-ray flux.   A similar
anticorrelation between optical flux and optical percentage polarization was
noted in P99; however, as noted by those authors, there is no analogous
anticorrelation  between radio polarization and either radio or optical flux
(their Figures 3-7 and \S\S 3-4).  

It is interesting to elaborate on this anticorrelation.  In three knot regions
-- HST-1, D-E and D-X (Figure 6, {\it top and left middle panels}) -- the X-ray
peak coincides exactly with the location of the optical polarization minimum,
while in two other regions, the X-ray flux peak is either slightly upstream
(knot E) or downstream (knot F) of the optical polarization minimum.  As has
already been noted by P99, several of these optical polarization minima tend to be
immediately downstream of regions of increased optical  polarization where the
magnetic field is perpendicular to the jet.    The morphology of the magnetic
field is not the same for all the inner jet  knots, however.  For example, in
knots HST-1, D-E and F (Figure 6, {\it middle panels}) large rotations of the
magnetic field  direction are seen just upstream of the X-ray flux maximum,
while much less rotation is seen in knot E (Figure 6, {\it right middle
panel}), where the  magnetic field appears to remain more closely parallel to
the jet direction.

Three X-ray bright knots in the inner jet are not strongly anticorrelated with
optical polarization.  The X-ray peaks of knots D-W and I (Figure 6, {\it left
middle} and {\it right middle} panels respectively) correspond to
regions of appreciable $(15-25 $\%)  but relatively constant  optical
polarization.  Both show some rotation of the magnetic field direction, but
again the details are different.  In knot D-W the magnetic field becomes nearly
perpendicular to the jet direction {\it downstream} (rather than upstream, as
in knot D-E and others discussed above) of the flux maximum, while in knots D-X and
I the orientation of the magnetic field seems unrelated to the jet direction.

The details of the relationship between X-ray flux and optical polarization
appear significantly different in the outer jet (Figure 6, {\it top and bottom
panels}).  Both the X-ray and optical flux peak of knot A, the brightest knot,
occur well downstream of the peak in optical polarization noted by P99.  The
X-ray flux maximum is in a ``valley'' of reduced polarization between two high
optical polarization regions (called HOP-1 and HOP-2 by P99) which surround
both the X-ray and optical flux peaks.  However, the optical polarization at
the X-ray flux maximum is still considerable ($\approx 35$\%).  Moreover, as
can be seen in Figure 6 ({\it bottom}), the magnetic field vectors in knot A
are consistently perpendicular to the jet direction, a marked contrast to what
is seen near flux peaks in the inner jet.  Much lower optical polarizations
($\sim 15-20$\%) are seen in a band stretching for $0\farcs4$ between the
optical flux maxima of knots A and B.  The X-ray flux peak of knot B, which
coincides with optical knot B1,  is different still: this knot is located in a
region of fairly high optical polarization, with magnetic field vectors that
tend to be parallel to the jet direction. Given the faintness of the regions
between knots B1 and C in the X-rays, it is impossible to comment on the
relationship between X-ray flux and optical polarization there.  The X-ray flux
peak corresponding to knot C1 occurs just downstream from an optical polarization
minimum, in a region where the optical polarization is increasing with distance
from the nucleus  but has not yet reached maximum.  However, the terminal X-ray
flux peak corresponds to the region between optical knots C and G, and does
correspond to a low optical polarization.

Comparing Figure 6 with Figures 1 and 3-5, two points can be made.   First, the
knots where one sees a better correspondence between X-ray flux maxima and
optical percentage polarization minima (e.g., knots HST-1, D-E, F and A), tend
to be brighter in the X-rays and have considerably smaller $\alpha_{ox}$  than
the X-ray flux maxima that have weaker X-ray/optical polarization
anti-correlations (compare Figure 4, 5 and 6).   The second point is that like
the X-ray flux peaks, the optical percentage polarization minima tend to
coincide with lower values of $\alpha_o$ and $\alpha_{ox}$, and possibly larger
$\alpha_{ro}$ values.  In \S 4 we will analyze these features in the light of
our spectral modelling.

\section{Physical Interpretation}

Our reanalysis of the {\it Chandra} data confirms steep X-ray spectra
($\alpha_x > \alpha_r$), and is thus consistent with a synchrotron origin for
the X-ray jet emission.  However, here we have attempted a more thorough
discussion of the M87 jet's morphology and broadband (particularly 
optical-to-X-ray) spectral energy distribution than previous authors.  The
correlation between X-ray emission and optical polarization is also discussed. 
The addition of these elements allows us to address the emission mechanism and
physical conditions in more detail than was possible in previous papers.

% We do not allow for any bulk relativistic motion of the jet (Biretta et al.
% 1995, 1999).  Such motion affects the flux densities and break frequencies
% (e.g., Begelman, Blandford \& Rees 1984).  The observed {\bf spectra} of the
% knots should  be converted to the rest frame of the jet and the modelling {\bf
% applied} to the rest frame spectra.  However, our model is intended to account
% for the broadband {\it shape} of the spectra  of the knots, which is unaffected
% by bulk relativistic motion.

\subsection{Synchrotron Emission Models and the Jet SED}

We have fitted synchrotron spectral models to each pixel of the radio through
optical map cube described in \S 3.1, using programs written by C. Carilli and
J. P. Leahy (Carilli et al. 1991; Leahy 1991).  The purpose of this exercise
was not only to determine basic parameters such as synchrotron break frequency
and injection index (an exercise which had already been carried out with the
full-resolution optical data by P01a, who emphasize uncertainties in the
determination of $\nu_{break}$), but also to use the results of the fits
to {\it predict} the X-ray flux and spectral index at various X-ray
energies  at each pixel for each model. We emphasize that the 
{\it Chandra} data were not used in the computation of these maps.
In this way, we can determine the
degree to which particle acceleration is or is not necessary at each position
along the jet, and diagnose the loci and energy dependence of accelerated
particles, where required.  Only the smoothed radio-optical data were used for
model fits.  

The models we fitted to the data were:

(1) A Jaffe \& Perola (1973) model (hereafter JP).  In this model, a power law
spectrum with $n(E_{e}) \propto E_{e}^{-p}$ ($E_{e}$ is the electron energy)
is injected at $t=0$ and allowed to 
evolve,
taking into account losses to synchrotron radiation and/or inverse-Compton
scattering.  There is continuous isotropization of the pitch-angle distribution
of the electron population with time, but no further particle injection.  The
resulting spectrum is a power law at low energies with $\alpha=(p-1)/2$, and an
essentially exponential rollover above the synchrotron loss break frequency. 
In P01a, the JP model was still regarded as viable based only on simultaneous
fitting of the entire radio-X-ray spectrum of the three knots which had been
seen by {\it ROSAT} and {\it Einstein}.  Our procedure rejects this
model with very high confidence, as the exponential high-energy rollover
underpredicts the X-ray flux at every pixel by many orders of magnitude and,
moreover, its slope at X-ray energies is much larger than those observed. The
JP model will therefore not be mentioned further in this paper.

(2) A Kardashev-Pacholczyk model (Kardashev 1962, Pacholczyk 1970; hereafter
KP).  This model assumes the same initial conditions as those in the JP model,
but there is no pitch-angle scattering, so a high-energy ``tail'' of particles
with small pitch angles develops. The resulting spectrum is a broken power law,
with $\alpha=(p-1)/2$ at  low frequencies and $\alpha=(2p+1)/3$ at high
frequencies.  The frequency of the break moves to lower values with  increasing
time.   Since the spectral index of the radio knots (Biretta et al. 1991) is
$\alpha_r \approx 0.6$, then $p=2.2$.  If $p_x=p_r$ (an assumption we will
return to later), then the KP model predicts X-ray spectral indices of
$\alpha_x=1.75-1.9$ for the integrated emission of a 
volume containing both sites of initial acceleration and regions into 
which the particles have moved; this   range of values
is somewhat larger than observed (Table 2, Figure 4).  Constancy of
particle pitch angle is also somewhat implausible, in view of the likely
scattering of relativistic particles by hydromagnetic waves (e.g., Wentzel
1974) and motion of particles into regions of different magnetic field
direction.  However, we retain this model in our discussion as it does not
underpredict the fluxes by orders of magnitude (see below).

(3) A ``continuous injection'' (hereafter CI) model (Kardashev 1962; Ginzburg
\& Syrovatskii 1968; Heavens \& Meisenheimer 1987, Meisenheimer et al. 1989). 
In this case, a power-law distribution of relativistic particles is
continuously injected at a constant rate.  When the entire source is included
in the telescope beam, there is a break of $\Delta \alpha = 0.5$ between the
low and high frequency spectra, which are both power laws ($\alpha=(p-1)/2$ at 
low frequencies and $\alpha=p/2$ at high frequencies).  If $p_x=p_r$, we expect
typical X-ray spectral indices $\alpha_x \approx 1.1$, somewhat smaller than
observed (Table 2, Figure 4).  P01a showed that such a CI model
cannot apply to knots A, B and D, since it overpredicts their X-ray fluxes by
factors of 8-50. 
%as if their  integrated radio-optical fluxes  were 
%described by such a  model, their X-ray fluxes were overpredicted by factors of 
%8-50.  
However, that result gave no indication of whether or not the CI model could
apply with more complex assumptions about the sites, or spectrum, of 
accelerated particles. 

We show in Figure 7 the results of a CI model fit,  for the purpose of
illustrating the trends in fitted break frequency $\nu_{break}$ and injection index
$\alpha_{in}$.   We stress that only the radio and optical data were used
to develop this model.  Note that the trends in  $\nu_{break}$ and
$\alpha_{in}$ are identical for the KP model, except for an offset in
$\nu_{break}$. 

Two interesting things can be seen in Figure 7.  First, the X-ray flux maxima
correlate very well with the highest-$\nu_{break}$ regions of the M87 jet.  
Thus regions of high $\nu_{break}$ are also anti-correlated with optical
polarization.   This indicates that the X-ray flux ``knows about''
$\nu_{break}$; therefore, the X-ray emitting particles cannot be an  entirely
independent population from those at lower energies. Instead, the correlation
indicates that the X-ray emission is best explained as the high-energy
extension of the radio-optical spectrum. Second, Figure 7 shows small
variations in $\alpha_{in}$.  At some level, this is a feature induced by the
data used in our modeling process, which included only a single radio point. 
This limitation forces $\alpha_{in} \simeq \alpha_{ro}$.
We have not included a second radio point in the modeling because a
radio spectral index map with sufficient resolution is not easily obtained;
moreover, comparing the Biretta et al. (1991) values of $\alpha_r$ with our
values of $\alpha_{ro}$ reveals evidence of significant spectral curvature
(typical $\Delta \alpha \approx 0.1-0.2$) between 
15 GHz and the near-infrared point at 2.05 $\mu$m.  We shall
nevertheless assume an injection index, $\alpha_{in}$, equal to the  local
value of $\alpha_{ro}$ at all energies in view of the small difference between
$\alpha_r$ and $\alpha_{ro}$.

We have used these spectral models to predict X-ray fluxes and spectral 
indices for the KP and CI models at several energies within the 
{\it Chandra}
band:  0.3, 0.5, 1, 2, 3, 5 and 8 keV.  This allows us to test the
applicability of each synchrotron model to the M87 jet.   

In Figure 8, we show the predictions at 1 keV for  the CI ({\it left panels})
and KP ({\it right panels}) models. 
For each model, we show the ratio of predicted 1 keV
flux to observed 1 keV  flux (i.e., $F_{pred}/F_{obs}$, {\it upper panels}),
and predicted spectral index at 1 keV ({\it lower panels}).  The spectral 
indices are essentially $\alpha_x=p/2$ (CI) or $\alpha_x=(2p+1)/3$ (KP).  
The models shown
in Figure 8 are for the case where we adopted $\alpha_{in}$ values from
fitting the radio-optical data (see above), 
and thus approximate very closely the values of $\alpha_{ro}$.
%The color scale for $F_{pred}/F_{obs}$ runs from 0.01
%(red) to 0.4 (blue) for the KP model and 1 (red) to 20 (blue) for the CI model, 
%while that for $\alpha_{x,pred}$ runs from 1.15 (red) to 1.25 (blue) for the CI 
%model and 1.75 (red) to 1.9 (blue) for  the KP model.

As can be seen, the  KP model yields  values of  $F_{pred}/F_{obs}$ generally
running between $0.10-1$ in knot regions and $0.01-0.1$ outside of knots.
Further, as noted above,  the KP model does not predict the  correct X-ray
spectral indices:   $\alpha_x$  is predicted to be $1.75-1.9$, compared to the
observed values of $1.3-1.6$.  Thus by 10 keV, the model underpredicts the
observed X-ray fluxes by factors of $3-500$, while below 1 keV it overpredicts
the observed X-ray fluxes in some knots. 

The CI model, by comparison, tends to consistently overpredict the 1 keV fluxes
(as originally noted by P01a).  As shown in Figure 8, we see generally lower
values of  $F_{pred}/F_{obs}$ in the inter-knot regions (often only 1--3), and
higher values in the knots (3--20).  Higher values are seen in
the region exterior to knot A than closer to the nucleus (this trend is also
seen in the KP model).  The CI model comes closer to the observed X-ray
spectral indices than the KP model,  predicting $\alpha_x \sim 1.15-1.25$,
somewhat smaller than the values we observe in most of the jet, although not
much flatter than the values seen in the nucleus, HST-1 and D.  We will discuss
below the possible reasons behind these differences.

\subsection{A Model for Particle Acceleration in the M87 Jet}

The previous section leaves us with an interesting dichotomy. Namely, 
throughout the jet, the CI model overpredicts the X-ray flux, while the KP
model underpredicts it.  A second aspect is that
the KP model predicts too steep an X-ray spectrum
throughout the jet,  while the CI
model predicts too flat a spectrum. 
We have already noted the physical
implausibility of the KP model.
Resolution of its problems in describing the observations
would require injection of high energy particles to increase the
X-ray flux and flatten the X-ray spectrum - i.e. the addition to the model of 
continuous injection at X-ray emitting energies. 
We thus feel that an appropriate approach is 
to modify the CI model in such a way as to remove the discrepancy with
observations.

The proposed model is thus a modified CI model in which it is assumed that:

(i) the volume within which
particle acceleration occurs is energy dependent,
being smaller for particles of higher energy, 

(ii) the spectrum of the
injected electrons is a power law $n(E_{e}) \propto E_{e}^{-p}$ below the
cut-off implied by (i), and 

(iii) the value of $p$ is independent of energy
and position.

We note that other modifications of the CI model can  be made to remove the
disagreement with observations and briefly address such possibilities in \S
4.4.   In the following, we derive the energy dependence of the volume within
which particles are accelerated.  For simplicity, we assume that electrons that
emit optical synchrotron emission are accelerated throughout the jet, i.e.,
with unit filling factor $f_{acc}$, even though in P99 we advanced a model of
partial energy stratification to explain the differences between the optical
and radio polarization morphology.  We can then derive $f_{acc}$ at higher
energies from:

$$f_{acc}(E_\gamma,x,y,z) = \Biggl({{F_{obs}(E_\gamma,x,y,z)}\over
{F_{pred, CI}(E_\gamma,x,y,z)}}\Biggr) , \eqno{(1)}$$ 

{\noindent{ where $E_\gamma$ is the energy of the photon emitted by the
electron.}

We expect $f_{acc}$ to be a function of both position (as $F_{pred}/F_{obs}$ is
observed to be; Figure 8) and energy.   Since, as shown in Figure 2, the {\it
Chandra} observation barely resolves the jet's width,  we feel it is most
appropriate to explore the variations of $f_{acc}(E_\gamma)$  as a function of
distance from the nucleus along the jet.  We show this in Figure 9 (top), which
shows $f_{acc}(r)$ at six different energies: $E_\gamma = $ 0.3, 1, 2, 3, 5 and
8 keV. 

Figure 9 (top) shows that at each energy $f_{acc}$ varies quite widely.   
We first discuss its variations as a
function of position, focussing on the 1 keV curve.
%Nearly all of %the jet has $f_{acc}<1$. The lone exception is the region between 
%$0\farcs5-2''$ from the nucleus, i.e., knot HST-1, which is also the only 
%region where $\alpha_{ox}$ is smaller than the predictions of the CI model.  
%Strongparticle injection {\it must} be taking place in HST-1, as it has a
%$\nu_{break}$ value 3-4 decades higher in frequency than the core (P01a,P01b), 
%in fact the highest in the entire jet (M02; see also Figure 5).  HST-1's high 
%value of $f_{acc}$ is almost certainly due to variability, as beginning in 
%1999-2000 it began to brighten dramatically, with its July 2000 optical flux 
%being more than a factor two higher than seen in 1998-1999 (Biretta et al. 
%2003).  It is likely that this component also increased in X-ray flux during 
%this time as well, although since no X-ray images were able to resolve HST-1 
%prior to 2000 it is impossible to say this for certain.  This brightening is 
%currently culminating in a series of at least three X-ray/optical flares during 
%2001-2003 (Harris et al. 2003, Perlman et al. 2003), and currently HST-1 is 
%$>10\times$ brighter in the optical than it was in 1998, and its X-ray flux 
%has also increased by a factor five over what we report here.   Fortunately, 
%no other region of the jet shows large amplitude
%variations, so $f_{acc}$ is not strongly affected by variability at other jet
%locations. Outside of the HST-1 region, we see that  
These spatial variations of
$f_{acc}$ at any given energy are highly significant, with  $f_{acc}$ 
generally declining with increasing distance 
from the nucleus.  Overall, in the inner jet one sees at 1 keV 
values of $f_{acc}=1-0.2$, i.e., particle
acceleration is taking place in  
100\% (in knots D and E) to 20\% (in knot I)
of the jet volume.
By comparison, in the outer jet we see values of $f_{acc} (1 ~{\rm keV}) =
0.2-0.07$, i.e., 
particle acceleration is taking place over 20\% (in knot A) to 7\% (in knot C)
of the jet volume. Thus, in general,
particle acceleration seems to be taking place in a larger
fraction of the jet volume in the inner jet than in the outer jet.  

Superposed on this general trend we also see variations on smaller angular
scales.   Not including the nucleus and HST-1, for which the values are
unreliable due to  variability, the highest value of $f_{acc}$ is seen at D-X. 
In the  inner jet, local minima in $f_{acc}$ appear close to the locations of
some of the knot flux maxima -- in particular knot D-E at $r=2\farcs8$ and knot
D-W at $r=3\farcs8$; however, there is no clear pattern to these variations.

One also notices in Figure 9 (top) an obvious energy dependence of $f_{acc}$,
in the sense that particle acceleration regions occupy a smaller fraction of
jet volume at higher energies.  This seems to  persist in all regions of the
jet (once again, the variations in energy at the positions of the nucleus and
HST-1  are not reliable due to variability).  As can be seen, the variation is
quite large -- at any given position in the jet, $f_{acc}$ varies by factors of
$3-20$ from the lowest (0.3 keV) to the highest (8 keV) energy.  One
simple parameterization of this energy dependence is a power
law: 

$$ f_{acc}(E_\gamma) \propto E_\gamma^{\xi} \propto E_e^{\xi/2}. \eqno{(2)}$$ 

{\noindent This
parameterization allows one to relate $\xi$ to the  parameters in the
synchrotron spectrum if the CI model holds.  In particular, since the CI model
predicts $F_{\nu, pred} = K_p \nu^{-p/2}$ (for an energy  independent
$f_{acc}$), and we observe $F_{\nu,obs}  = K_{obs} \nu^{-\alpha_x}$ (where
$K_p$ and $K_{obs}$ are constants), we have }

$$ f_{acc} = F_{\nu,obs}/F_{\nu,pred} = constant \times  \nu^{-(\alpha_x - p/2)}
\eqno{(3)}$$

$$ \therefore \xi = (-\alpha_x + p/2) = (-\alpha_x+\alpha_{in}+0.5)
\eqno{(4)}$$

We have calculated the value of $\xi$ for different locations along 
the jet using equation (4). The values thus obtained are in
approximate agreement with those found by fitting a power-law
to the energy dependence of $f$ (Figure 9, top).   
%With this in mind, we have fitted a power law at each position to the energy
%dependence shown in Figure 9 (top).  The result is shown in Figure 9 (bottom), 
%which shows the values of $\xi$:
%
%\begin{flushleft}
%
%\indent {(i) calculated from spectral modelling (giving $\alpha_x$) at 
%various locations along the jet,  using Eq. (4) [points plus error bars], and 
%
%(ii) calculated numerically using the energy depedence of $\xi$ 
%indicated by Figure 9 (top). } 
%
%\end{flushleft}
%
%{\noindent These two approaches are entirely equivalent and show good agreement except
%within $2''$ of the nucleus, where the results are compromised by variability,
%and between $12''$ and $14''$ from the nucleus.  
Figure 9 (bottom) shows that in the inner jet, $\xi = -0.4 \pm 0.2$.  
This small amount of variation is interesting, particularly
given how  much $f_{acc}$ varies along the jet.
More negative values ($\xi =-0.6$ to $-0.8$) are seen in two relatively isolated
locations, $3''$ and $8-9''$  from the nucleus.  
We also see $\xi=-0.7 \pm 0.2$ in knots A and B, $12''-15''$ from the nucleus.
%
%In the outer jet, between $12''$ and $14''$ from the nucleus, there is some
%discrepancy between the analytic and numeric calculations of $\xi$, with the
%analytic estimate indicating $\xi = -0.7 $ to $-0.8$ and the numerical estimate 
%indicating $\xi=-0.4$.  This difference may be related to the different 
%energy ranges in the calculations.}
%The last two regions
%where we see  $\xi < -0.5$ are at $>16''$ and $18''$ from the nucleus,
%corresponding respectively to the regions between knot B and C1, and C1 and
%C2.  Interestingly, in the knot C region we observe a flattening of the X-ray
%spectrum (Figure 2) but relatively constant $\alpha_{ox}$ (Figure 3).  
If one compares the top and bottom panels of Figure 9,
one notices a possible
correspondence between the first two of these
regions where we see the most negative exponents $\xi$ and 
small local values of $f_{acc}$ at low X-ray energies, 
suggesting that the decline in 
$f_{acc}$ towards higher X-ray energies represents a continuation of a decline
in $f_{acc}$ from optical to X-ray energies.  
A recent analysis of deep UV imaging
of the M87 jet with HST (Waters \& Zepf 2005) found evidence that CI models
with $f_{acc}$ = 1 fitted to the P01a radio-optical spectra exceed the
observed flux at 1700 \AA.  This result is also qualitatively consistent with a
decline in $f_{acc}$ from optical to higher energies, but may more simply
reflect uncertainty in fitting idealised models to fluxes measured at closely 
spaced wavelengths.

From the continuity of $f_{acc}$ and $\xi$ along the jet 
(Figure 9) we are forced to conclude that particle
injection and acceleration in the M87 jet must occur both within the knot
regions and outside them.   Jester et al. (2001, 2002) came to a similar
conclusion for the 3C 273 jet based solely on radio-optical data.   

\subsection{Sites and Types of Particle Acceleration}

The previous analysis established that in order to sustain the X-ray emission
in the M87 jet, {\it in situ} particle acceleration almost certainly occurs, 
both within knots and outside them.  Yet if the X-ray emission 
can be represented by the CI model, the volume within which X-ray emitting 
particles are accelerated cannot fill the entire jet volume at any location,
but instead fills only a fraction $f_{acc}$ which varies with both position
and energy.

%As already noted, it is a bit surprising that the curves shown in Figure 9a
%show somewhat lower (rather than higher) values of $f_{acc}$ within knots.  How
%can we explain this feature?  One possibility is that the knots represent
%shocks occurring within the jet, in which the flow is compressed.  Such a shock
%would be a natural place for enhanced particle acceleration to occur, and in
%such a case, one would see a decrease in $f_{acc}$ with the decrease being 
%inversely proportional to the compression ratio in the shock. As has already
%been noted (P99), several of the knots in the M87 jet also  show polarization
%signatures which are consistent with shocks.  It is therefore appropriate to
%revisit the comparison between the optical polarimetry and the X-ray image  of
%the M87 jet in  the light of this analysis.

As already noted in \S 3.3, there is a strong correlation between the locations
of optical polarization minima and X-ray flux maxima in several regions;
this correlation is seen in knots HST-1, D-E, D-X, F and possibly A and C1.
Figure 6 shows that some of the regions in the inner jet (i.e., HST-1, D-E, D-X
and F) show increased optical percentage polarization and perpendicular
magnetic field just upstream (by 0\farcs1 -- 0\farcs3) of the X-ray peaks, low
optical percentage polarization at the X-ray flux maxima, and then an increase
of the optical percentage polarization immediately downstream of the X-ray
peak.  Often the downstream magnetic field direction is parallel to the jet.  
Knot A is similar but here the downstream field is perpendicular to the jet.
Ongoing {\it HST} polarimetry of the jets of several other nearby radio
galaxies (Perlman et al., in prep.) shows that this trend of low optical
polarization at X-ray flux maxima, often accompanied by changes in the
polarization PA, seems to persist in the population of X-ray synchrotron
emitting jets as a whole. 

Particle acceleration at shocks (e.g., Blandford \& Ostriker 1978) through the
first-order Fermi process is generally believed to occur in jets. 
For ultrarelativistic shocks, this model
predicts an injection index of $p=2.23$ (Kirk 2002; Kirk \& Dendy
2001), in very good agreement with our analysis of these data.
Further, the low observed polarization at X-ray peaks would result from
beam averaging over the pre- and post-shock regions, since the field direction
is expected to be different in these two regions.  The perpendicular field
upstream of the knots might result from a second, transverse upstream shock. 
Alternatively, changes in polarization direction may result from differing 
contributions from a perpendicular field region in the center of the jet and a
parallel field in the sheath (e.g., P99). 

Our model implies that the volume within which particles are accelerated
decreases with increasing particle energy. In other words, there is an
upper cut-off to the energy of accelerated particles in any given element of
volume. In the context of the first-order Fermi process in non-relativistic
shock waves (e.g., in supernova remnants), an
upper cut-off may occur if there are no waves (because of ion-neutral
damping) ahead of the
shock capable of reflecting particles above a certain energy; such particles
are then 
not reflected back and forth across the shock, as is needed to gain energy
(Drury, Duffy \& Kirk 1996). In the ultra-relativistic case,
damping of such long 
wavelength modes
would probably only be important downstream of the shock 
(Kirk, private communication). 
One also expects a high energy cut-off at energies where the
acceleration timescale is equal to the cooling time to synchrotron or
inverse Compton losses. It is thus plausible that a high energy cut-off 
should be present, but details
are beyond the scope of this paper.

\subsection{Alternative Models}

The model discussed in \S 4.2 is not unique. 
We see at least three alternative possibilities 
and deal with each of these briefly in
this subsection.

One possible alternative model is that the index of the electron energy
spectrum at injection
is larger at X-ray emitting energies than at lower energies. 
To
account for the typical X-ray spectral index, $\alpha_x \approx 1.4$, would
require an index at injection of
$p_x=2.8$, in contrast with $p_r \simeq p_{ro} \simeq p_o = 2.2$ at
lower energies. This is, of course, an {\it ad hoc} postulate, but no more
{\it ad hoc} than our suggestion that the volume filling factor within which
particles are accelerated,  $f_{acc}$, declines with increasing energy at X-ray
energies. These two pictures are very similar in that they
represent simple phenomenological ways of modifying the classical CI model
to allow agreement with the observations. However, we prefer our model in
which $f_{acc}$ declines with increasing energy for two reasons. First, 
the index at injection is $p=2.2$ at all energies and all locations along
the jet, consistent with current theoretical results for ultrarelativistic
shocks (Kirk \& Dendy 2001; Kirk 2002). Second, in the model in which the
electron energy
spectrum at injection
is steeper at X-ray emitting energies than at lower energies, the value of
$p_x$ must vary along the jet because $\alpha_{x}$ varies.

 Another possible model is 
that the X-rays may come from a different population of relativistic electrons
from those responsible for the optical emission.  This could be
related to the ``stratified flow'' proposed by P99.   
This model seems less likely to us because we do not observe $\alpha_x <
\alpha_{ox}$ or $\alpha_{ox} < \alpha_o$ in any region of the M87 jet (contrary
to the analysis of WY02; see \S 3.2 and also WY02's erratum [Wilson \& Yang
2004]). Jester et al. (2002) have suggested a model of this type for some
regions of the 3C 273 jet; however those 3C 273 jet components {\it do show}
definite spectral hardenings in the ultraviolet, unlike M87.  

Finally,  within the classical CI model, it is possible for $\alpha_x -
\alpha_r$ to equal some value other than 0.5.  Such can occur  in relativistic
electron diffusion loss models which allow the magnetic field and diffusion
coefficient to vary as a function of distance from the location where the
electrons are injected and also allow the diffusion coefficient to be
energy  dependent (Wilson 1975).  Also, Coleman \& Bicknell (1988) 
investigated models in which relativistic particles are accelerated in
adiabatic, non-relativistic bow shocks, and the electron distribution function
evolves downstream through both adiabatic and radiative (synchrotron) losses. 
They too found breaks larger than 0.5.  Unfortunately little is known about
particle propagation and magnetic field variations within the M87 jet, so 
models of this type are beyond the scope of this paper.

\section {Summary}

We have performed a re-analysis of the deep {\it Chandra} image of  the M87
jet, first analyzed by WY02.  This analysis improves on that of WY02 in several
respects, including an improved instrumental calibration  as well as image
deconvolution.  The former has allowed us to obtain more reliable X-ray spectra
along the jet, while the  latter allowed us to improve the spatial resolution
by nearly 50\%, and fully separate knots HST-1 and I from adjacent emission for
the first time.  There is  evidence for slight spectral variations $\Delta
\alpha \sim 0.3$ along the jet, with the flattest spectra ($\alpha_x = 1.3$)
observed in knots HST-1, D and C, and somewhat steeper spectra ($\alpha_x =
1.6$) in knots F, A and B.   A careful comparison of the {\it Chandra} data to
the multiwaveband {\it HST} imaging and polarimetry data of P99 and P01a has
been performed in  order to analyze the broadband spectrum of the jet, and
diagnose loci and mechanisms of particle acceleration.

{\it In situ} particle acceleration almost certainly occurs within the M87
jet.  This has been demonstrated not only from particle lifetime arguments, but
also from spectral fits to the broadband spectra. We have used the absolute
fluxes and spectra throughout the X-ray band to determine the volume filling
factor, $f_{acc}$,  of regions within which particles are accelerated in the
M87 jet.  A continuous injection (CI) model in which both  $f_{acc}$ and the
power-law index of the energy spectrum of the injected electrons are constant
and independent of energy predicts $\Delta \alpha = 0.5$, where $\Delta \alpha$
is the change in spectral index between radio (or radio to optical) and X-ray
frequencies.  In contrast, we observe $\Delta \alpha = 0.7-1.0$, in agreement
with the conclusion that a CI model with $f_{acc}=1$ overpredicts the X-ray
emission by large factors (P01a, M02, WY02).  To account for this larger
$\Delta \alpha$, we have developed a model in which the filling factor
$f_{acc}$  varies as a function of position and energy from  $\sim 1.0-0.01$ at
1 keV, with a general decline as a function of both increasing distance from
the core and increasing particle  energy.  Describing the energy dependence of
the filling factor by  $f_{acc} (E_\gamma) \propto E_\gamma^\xi \propto
E_e^{\xi/2}$ (where  $E_\gamma$ is the photon energy and $E_e$ is the energy of
the radiating electron), we find $\xi= -0.4 \pm 0.2$ in most of the inner jet
and $\xi = -0.7 \pm 0.2$ for knots A and B. In this model, the index, $p$, of
the relativistic electron energy spectrum at injection is $p=2.2$ at all
energies and locations, in excellent agreement with the predictions of models
of cosmic ray acceleration by ultrarelativistic shocks.  

The X-ray peaks in the jet often coincide with minima in the optical percentage
polarization, i.e., regions where the magnetic field is not ordered.  We have
suggested that this effect results from shock waves at the X-ray peaks.  The
shocks both accelerate X-ray emitting electrons and reorient the field,
resulting in low polarization through beam averaging. A tendency for the field
to align perpendicular to the jet upstream of the X-ray peaks may reflect a 
second, transverse shock.

The  need for high energy particle acceleration (X-ray emission
requires $\gamma \sim 10^7 - 10^8$) in the M87 jet is confirmed by the
{\it Chandra} data; 
however, significant uncertainties remain, which would be
alleviated by deeper, higher-resolution X-ray data.  A longer integration would
determine the variations in X-ray spectral index with greater accuracy.  We
intend to revisit this issue in a future paper by adding together the multiple
{\it Chandra} observations of M87 (Cheung et al., in prep.), which by now
amount to well over 200ks.  The case for better resolution is equally clear: 
we do not resolve the inner jet well in the transverse direction,  though
we do resolve knots A, B and C in the transverse direction. Observations with
higher angular resolution would allow us to better pin down the relationship
between X-ray flux, spectral index and optical properties, as well as offsets
between component maxima in different bands.

\begin{acknowledgments}

We wish to thank M. Karovska, A. Siemiginowska, D. P. Huenemoerder and A. Ptak
for ideas regarding the deconvolution of {\it Chandra} data.  We acknowledge 
J. A. Biretta,
M. Birkinshaw, C. C. Cheung, D. E. Harris, M. Georganopoulos, S. Jester,
J. G. Kirk, A. P. Marscher, C. P. O'Dea, W. B. Sparks and D. M. Worrall 
for interesting
conversations.  We thank an anonymous referee for comments that improved this
paper.  ESP acknowledges support from NASA under LTSA grant NAG5-9997, as well
as {\it HST} grants STGO-7866.01, STGO-09142.01, STGO-09705.01, STGO-09847.01
and {\it Chandra} grant SAO-03700786.  ASW thanks NASA for support under {\it
Chandra} grants NAG-81027, NAG-81755 and G023147X and under LTSA grant
NAG5-13065.

\end{acknowledgments}

\vfill\eject

\begin{figure}

\centerline{\includegraphics*[scale=0.7]{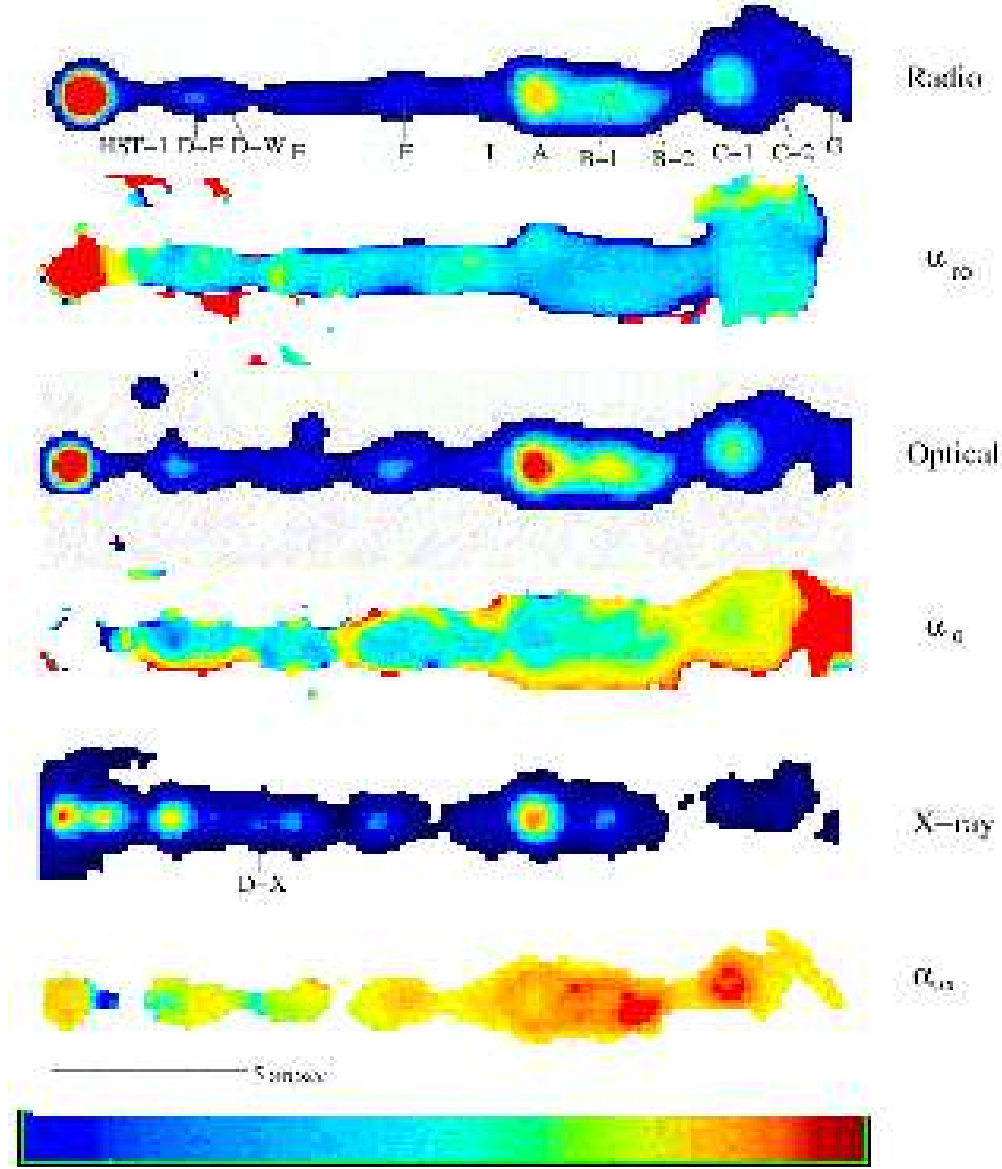}}

\caption[]{Images of the M87 jet. 
{\it Top panel.} Radio (VLA 15 GHz).
{\it Second panel.} Radio-optical spectral index. 
{\it Third panel.}  Optical (HST F814W).
{\it Fourth panel.}  Optical spectral index.
{\it Fifth panel.}  X-rays ({\it Chandra} 0.3-1.5 keV).
{\it Sixth panel.}  Optical-X-ray spectral index.  
The {\it Chandra} image has been processed so that pileup is unimportant, and
deconvolved using AIPS VTESS as described in \S 2.1, and galaxy light has been
subtracted from the optical  image as described in P01.  The radio and optical
images and spectral index maps were smoothed with Gaussians to a resolution of
$0\farcs5$ to match the FWHM of the PSF-deconvolved {\it Chandra} image (see
\S\S 2, 3).  The color scales of the spectral index images are $\alpha_{ro}$:
0.85 (red) to 0.6 (blue), $\alpha_o$:  1.5 (red) to 0.4 (blue), and
$\alpha_{ox}$: 1.6(red) to 0.9 (blue).  All images and maps have been rotated
so that the jet, which is at  PA $-69.5^\circ$ (North through East) is along
the x-axis.  The optical image suffers from saturation at the nucleus. The
images  have been registered by aligning the images of the nuclei, which can
be seen at far left.  See \S 3 for discussion.}

\end{figure}

\vfill\eject

\begin{figure}

\centerline{\includegraphics*[scale=0.6]{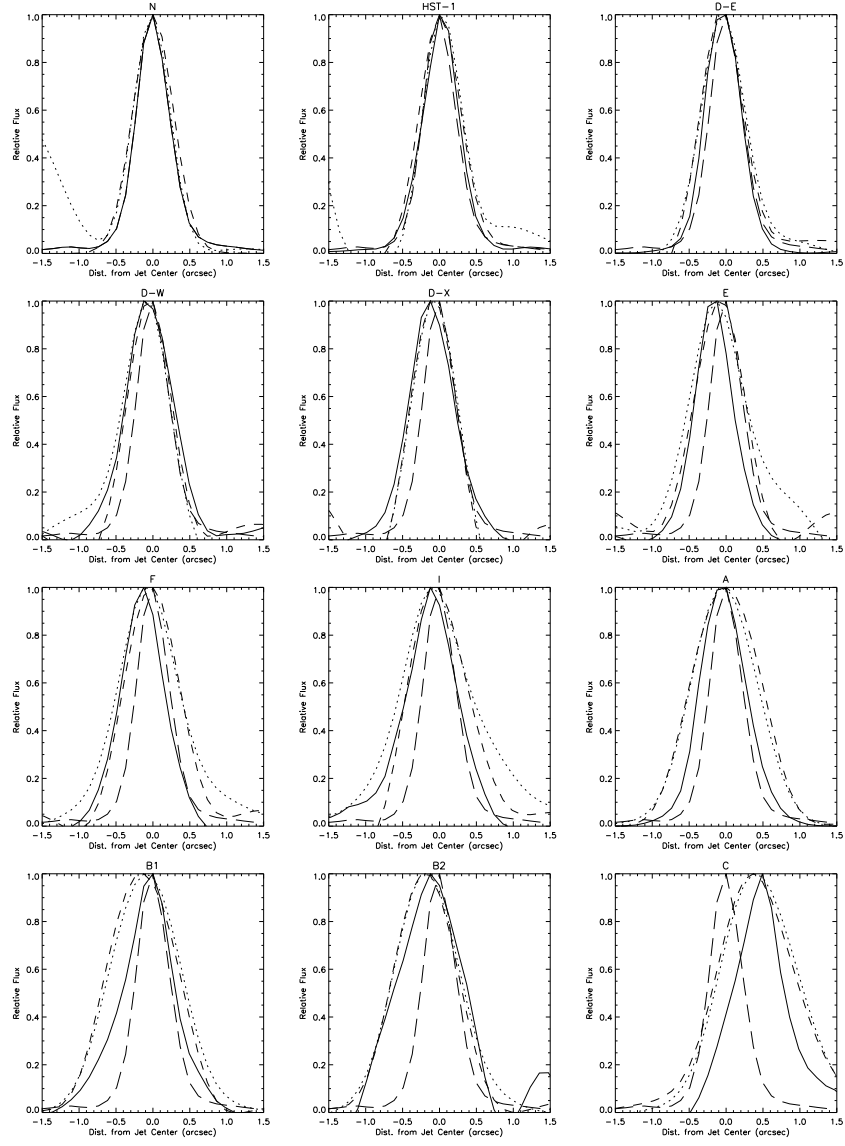}}

\caption[]{Plots of the profile of various jet components in the direction
perpendicular to the jet, in the X-ray  (solid line), radio (dotted line) and
optical (dashed line).  For comparison, in each panel (except that for the 
nucleus) we have plotted the {\it Chandra} PSF, as represented by the  profile
of the nucleus (long-dashed line).  All panels are centered at the jet
centerline, so that the shifts seen represent real offsets north (positive
values) or south of the jet centerline.  Knots A, B and C are clearly
resolved in the {\it Chandra} data, in the direction perpendicular to the jet. 
A few other regions are possibly resolved; however, we do not resolve most of
the inner jet components.}

\end{figure}
\vfill\eject

\begin{figure}

\centerline{\includegraphics*[scale=0.75]{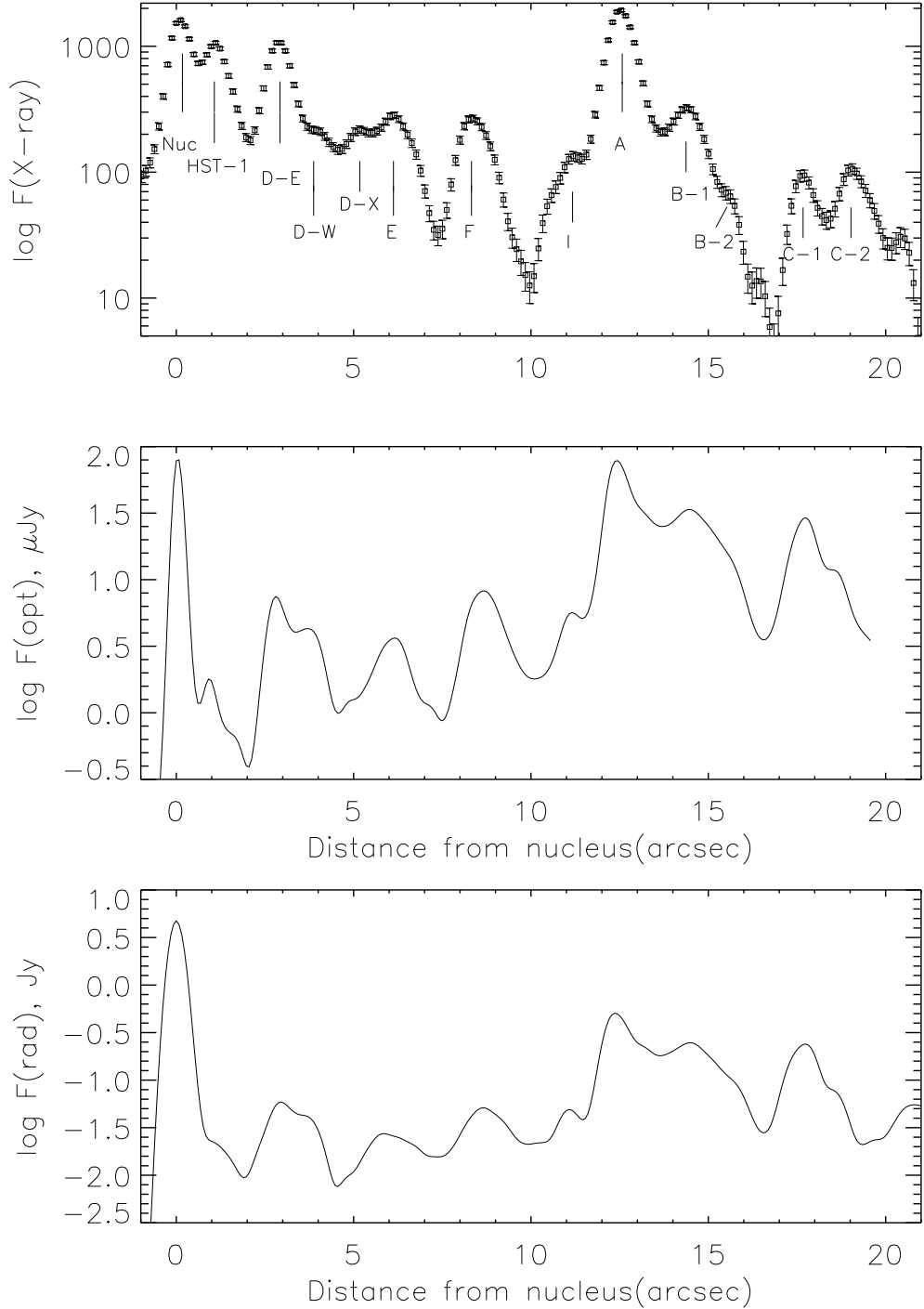}}

\caption[]{Runs of flux in the X-ray (0.3-1.5 keV band, at top), Optical (HST
F814W, middle), and Radio (VLA 15 GHz, at bottom) bands.  Flux in all three
panels is plotted on a logarithmic scale.  The optical and radio panels have
been smoothed with $0.4''$ Gaussians to match the  resolution of the {\it
Chandra} data.  See \S 3 for discussion.}

\end{figure}
\vfill\eject

\begin{figure}

\centerline{\includegraphics*[scale=0.75]{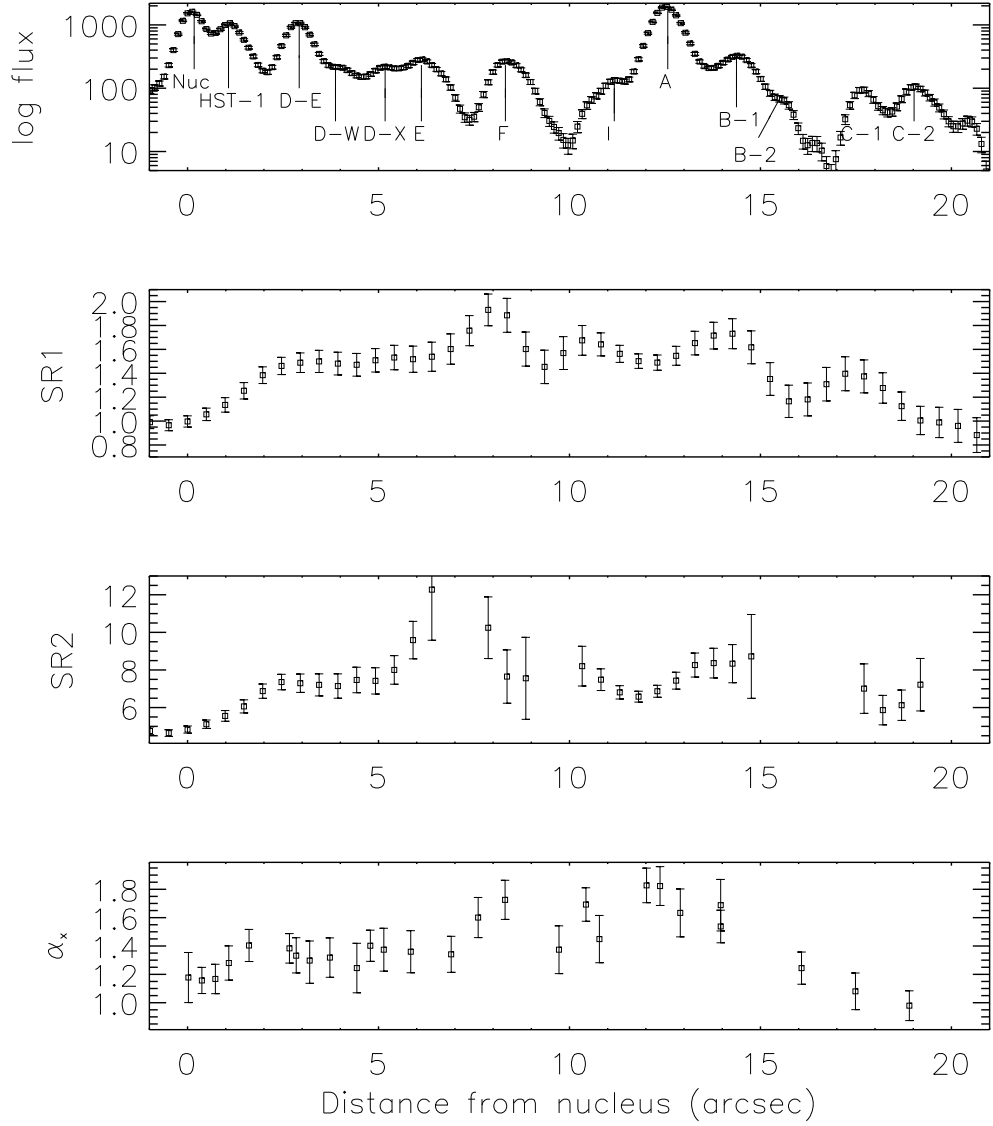}}

\caption[]{ Runs of flux and X-ray spectral information along the jet. 
{\it Top panel.}  Total X-ray flux.
{\it Second panel.} X-ray softness ratio $SR1$ = F(0.3-1 keV) / F(1-3 keV).
{\it Third panel.} X-ray softness ratio $SR2$ = F(1-3 keV) / F(3-10 keV). 
{\it Fourth panel.}  X-ray spectral index $\alpha_x$.  
Small variations are seen in the X-ray spectrum of the jet,
with flatter spectra seen in the nucleus and knots HST-1, D and C, and steeper
spectra in knots F, A and B.  See \S 3.2 for discussion.}

\end{figure}
\vfill\eject

\begin{figure}

\centerline{\includegraphics*[scale=0.65]{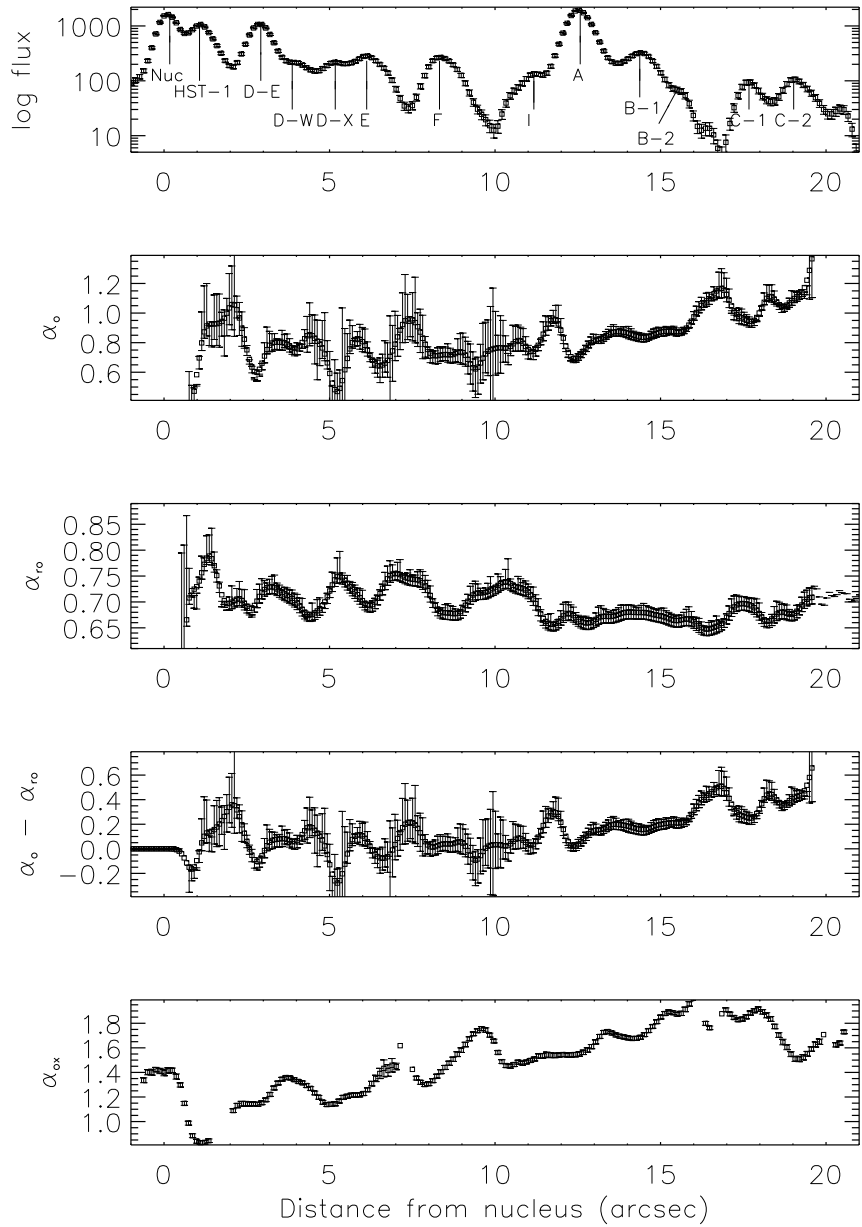}}

\caption[]{ The run of X-ray flux compared to broadband spectral indices.
{\it Top panel.} X-ray flux.
{\it Second panel.} Optical spectral index $\alpha_o$. 
{\it Third panel.} Radio-optical spectral index $\alpha_{ro}$ 
{\it Fourth panel.}  $\alpha_o-\alpha_{ro}$
{\it Fifth panel.} Optical-X-ray spectral index $\alpha_{ox}$. 
Values of $\alpha_o$,  $\alpha_{ro}$ and 
$\alpha_o-\alpha_{ro}$ are not shown within $0\farcs3$ of the nucleus due to
the saturation of the optical images at the position of the nucleus.  
The $\alpha_o$,  $\alpha_{ro}$ and $\alpha_o-\alpha_{ro}$ runs have been
smoothed with a Gaussian to a resolution of $0\farcs5$.   
See \S \S 3.2, 4.1 for
discussion.}

\end{figure}
\vfill\eject

\begin{figure}

\centerline{\includegraphics*[scale=0.7]{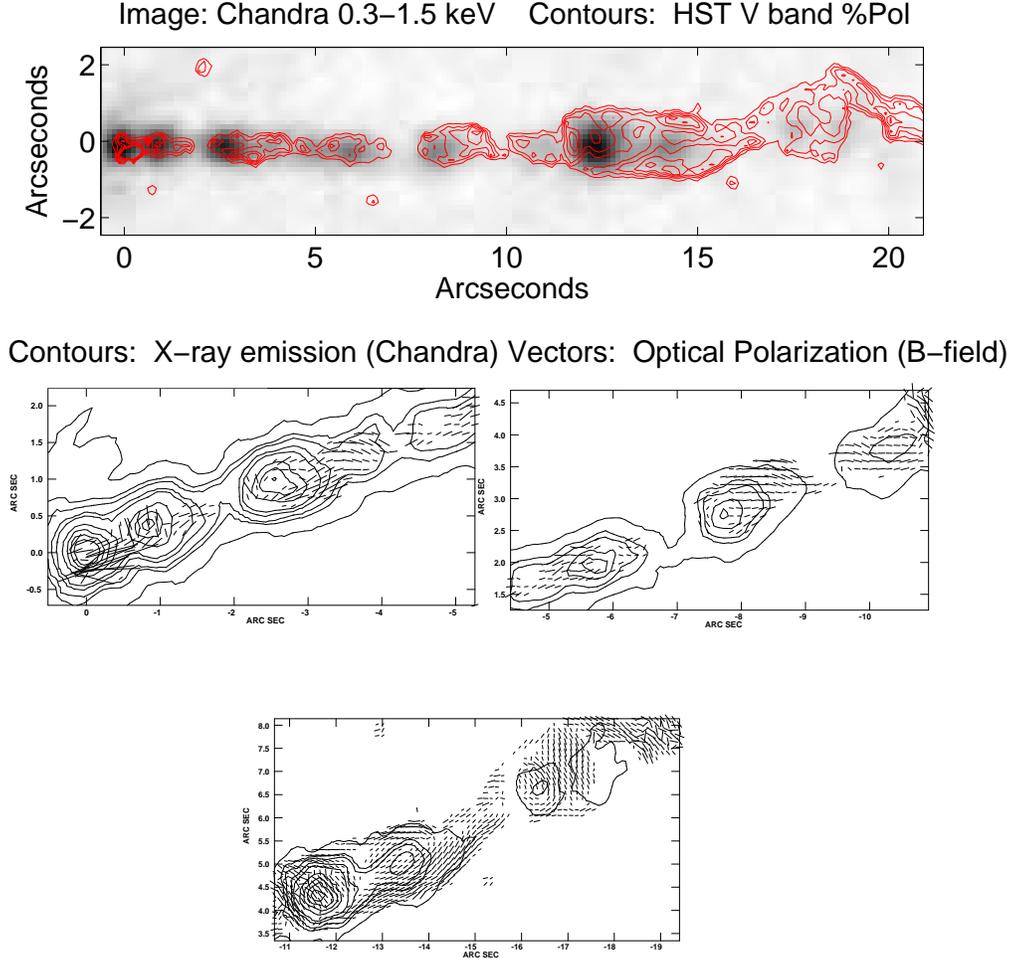}}

\caption[]{ Plots comparing X-ray flux to optical polarization.  At top, we
show the {\it Chandra} 0.3-1.5 keV image in greyscale, with red contours
representing the level of polarization in the optical (in percent).   A
logarithmic scale was used for the greyscale of the {\it Chandra} image, and
contours are shown at 5, 15, 25, 35, 45, 55 and 65\% polarization. The
bottom three images show the X-ray flux in contours, with optical polarization
({\bf B} field) vectors overplotted.  A vector $1''$ long corresponds to 200\%
polarization.  The optical polarization images shown here have not been smoothed
with a $0\farcs5$ Gaussian (unlike the total flux images discussed and shown
previously), in order to bring out details in the magnetic field configuration
of the jet.  As can be seen, there is excellent agreement between the positions
of X-ray flux maxima and optical polarization minima, particularly in the X-ray
brightest knots.  See \S \S 3.3, 4.2 for discussion.}

\end{figure}
\vfill\eject

\begin{figure}

\centerline{\includegraphics*[scale=0.6]{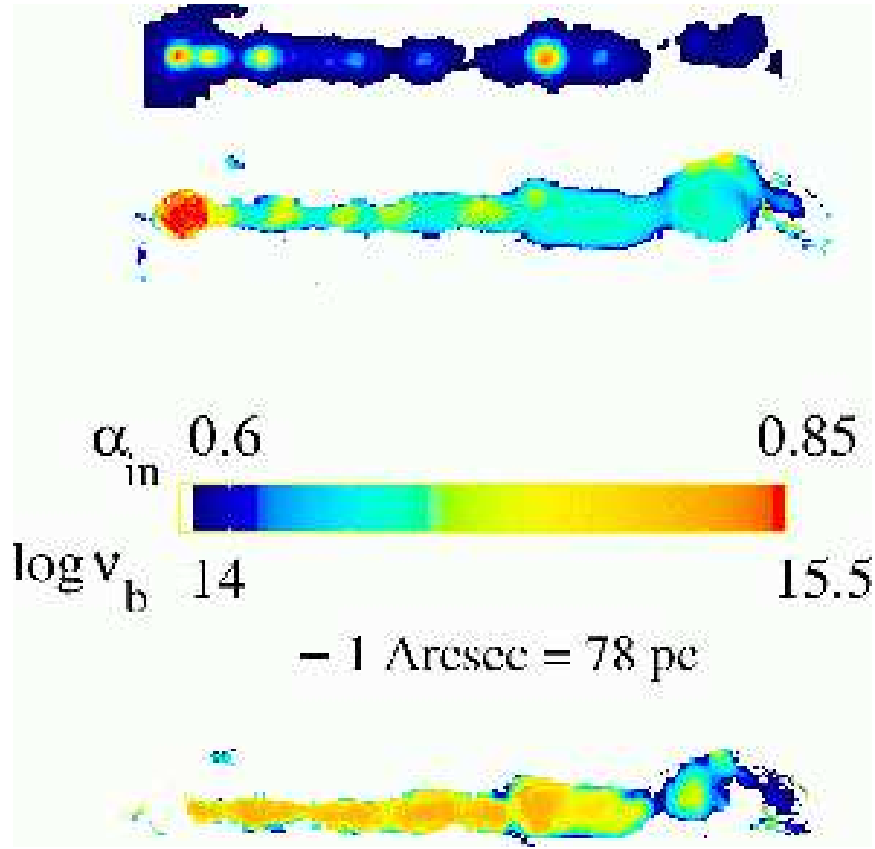}}

\caption[]{The results of modeling the radio-optical broadband spectrum with a
continuous injection (CI) synchrotron emission model in which the volume
filling factor of regions of particle acceleration is energy independent (so
$\Delta \alpha = 0.5$).  The top panel shows the deconvolved X-ray image of the
M87 jet, for reference.  The middle panel shows injection index $\alpha_{in}$,
with the color scale running from 0.85 (red) to 0.6 (blue).  At bottom, we show
the synchrotron break frequency $\nu_{break}$, with a color scale that runs
from $10^{14}$ Hz (blue) to $10^{15.5}$ Hz (red). All data used for these maps
has been smoothed with a   $0\farcs5$ Gaussian prior to modeling; full
resolution maps of these quantities have appeared previously in P01.  The
trends shown in the KP model image of $\nu_{break}$ are very similar, with
essentially only a zero-point offset representing the overall differences in
model characteristics (as described in Leahy 1991 and Carilli et al. 1991). The
KP and CI models give identical $\alpha_{in}$ images, as expected since that
value is constrained by only the low frequency spectral index.  As shown, the
X-ray flux maxima correspond well with regions of high $\nu_{break}$, but there
is no correlation between X-ray flux and $\alpha_{in}$. See \S 4.1 for
discussion.}

\end{figure}
\vfill\eject

\begin{figure}

\centerline{\includegraphics*[scale=0.75]{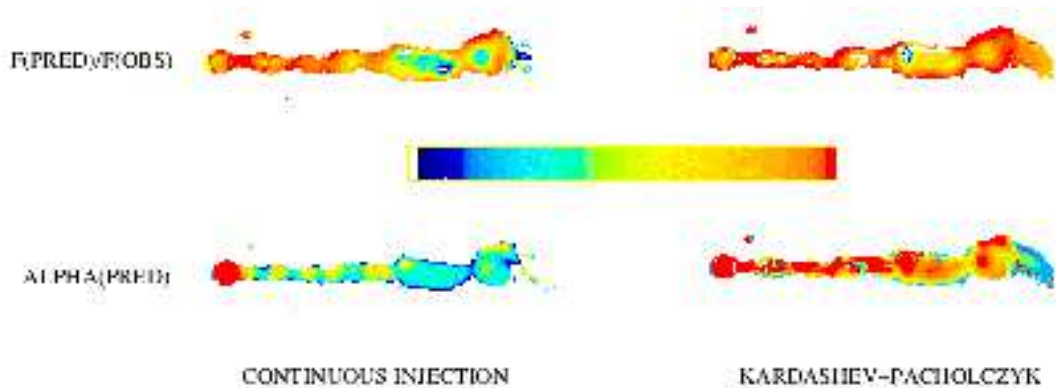}}

\caption[]{The  distributions of the
ratio $F_{pred}/F_{obs}$ and the predicted spectral index at X-ray 
frequencies 
for the continuous injection (CI,
left)  and Kardashev-Pacholczyk (KP, right) synchrotron models, with energy
independent filling factor $f_{acc}$ for regions of particle acceleration. The
top two panels show the ratio $F_{pred}/F_{obs}$ at 1 keV, with color scales
that run from 1 (red) to 20 (blue) for the CI model and 0.01 (red) to 0.4
(blue) for the KP model.  The bottom two panels show the predicted spectral
index at frequencies $\nu >> \nu_{break}$  (computed at 1 keV), with color
scales that run from 1.15 (red)  to 1.25 (blue) for the CI model and 1.75 (red)
to 1.9 (blue) for the KP model.} 

\end{figure}
\vfill\eject

\begin{figure}

\centerline{\includegraphics*[scale=0.6]{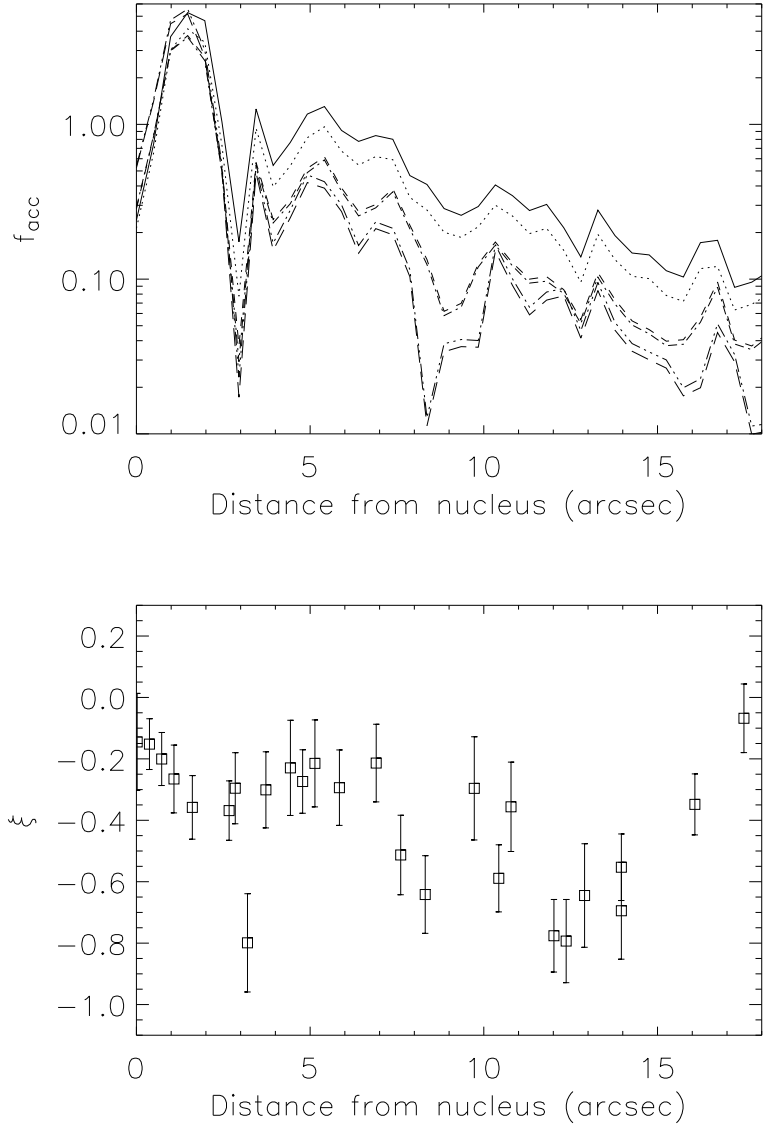}}

\caption[]{At top, we plot the filling factor,  
$f_{acc}$, of the regions within which particles radiating X-ray
synchrotron radiation at $E_\gamma$ are accelerated, versus
distance from the nucleus, for six 
different values of $E_\gamma$.  From top
to bottom, these are: 0.3 keV (solid line), 1 keV (dotted line), 2 keV
(short-dashed line),  3 keV (dot-dashed line), 5 keV  (dot-dot-dot-dashed line)
and 8 keV (long-dashed line).  The high values at $r=0\farcs5-2''$ result
from massive variability during 1999-2003 in knot HST-1, and should not
be taken as representative of the knot at a single epoch.
The general decrease in $f_{acc}$ with increasing $r$ is notable.  
At bottom, we plot the injection exponent $\xi$
($f_{acc}(E) \propto E^\xi$) versus distance from the nucleus, computed
using equation (4). See \S 4.3 for discussion.

}
\end{figure}
\vfill\eject

\begin{deluxetable*}{cllcclllccl} 
\tablecolumns{11}
\tablewidth{0pt}
\tablecaption{X$-$ray Positions and Sizes of Jet Components}

\tablehead{
\colhead{(1)}&\colhead{(2)}&\colhead{(3)}&\colhead{(4)}&\colhead{(5)}& \colhead{(6)} & \colhead{(7)} & \colhead{(8)} & \colhead{(9)} & \colhead{(10)} & \colhead{(11)}\\
\colhead{}&\multicolumn{2}{c}{X$-$ray Position
(J2000.0)}&\multicolumn{2}{c}{Comp Size (FWHM)} & Distance from & \multicolumn{2}{c}{Optical Position(J2000.0)}\\
\colhead{Component}&\colhead{RA}&\colhead{Dec}&\colhead{Maj}&\colhead{Min}& \colhead{Nucleus ($r$)} & \colhead{RA} & \colhead{Dec} & \colhead{DeltaRA} & \colhead{Delta Dec} & \colhead{Notes} \\
\colhead{}& \colhead{h m s} & \colhead{$^\circ ~ ' ~ ''$} & \colhead{arcsec}&\colhead{arcsec}& \colhead{arcsec}& \colhead{h m s}&\colhead{$^\circ ~ ' ~ ''$}& \colhead{arcsec}& \colhead{arcsec} 
}
%Comp	RA(Xray)	Dec(Xray)	MAJ	MIN	$r$		OPT RA		OPT DEC		DELTA RA	DELTA DEC	NOTES

\startdata
Nucleus	&12 30 49.417	&12 23 28.02 	&0.70	&0.58	&0		&12 30 49.417	&12 23 28.02	&0		&0		& a\\
	&(0.001)	&(0.01)		&(0.04) &(0.03) & 		&(0.002)	&(0.02)&&\\
HST$-$1	&12 30 49.355	&12 23 28.32	&0.79	&0.60	&0.96		&12 30 49.363	&12 23 28.42	&+0.12		&+0.11	&b\\
	&(0.002)	&(0.02)		&(0.06) &(0.04) &(0.04)	&	(0.001)&	(0.01)&		(0.03)&		(0.03)&\\
D$-$E	&12 30 49.239	&12 23 28.92	&0.85	&0.73	&2.77		&12 30 49.233 &	12 23 29.00&	$-$0.08&		+0.08	&	\\		
	&(0.002)	&(0.02)		&(0.06) &(0.05)	&(0.04)	&	(0.002)&	(0.02)&		(0.03)&		(0.03)&\\
D$-$W	&12 30 49.165	&12 23 29.34	&0.49	&0.49	&3.92		&12 30 49.188	&12 23 29.18	&+0.33		&$-$0.16		&c\\
	&(0.003)	&(0.04)		&(0)	&(0)	&(0.06)	&	(0.005)&	(0.05)&		(0.06)&		(0.06)&\\
D$-$X	&12 30 49.089	&12 23 29.61	&0.49	&0.49	&5.05		&12 30 49.070	&12 23 29.82&	$-$0.27		&+0.21		&c,d\\
	&(0.003)	&(0.04)		&(0)	&(0)	&(0.06)	&	(0.003)&	(0.04)&		(0.06)&		(0.06)&\\
E	&12 30 49.030	&12 23 29.94	&1.4	&0.9	&6.00		&12 30 49.017	&12 23 30.10	&$-$0.18		&+0.16		&d\\
	&(0.002)	&(0.03)		&(0.2)  & (0.1)	&(0.05)	&	(0.006)&	(0.06)&		(0.07)&		(0.07)&\\
F	&12 30 48.884	&12 23 30.74	&1.6	&1.2	&8.27		&12 30 48.862	&12 23 31.00&	$-$0.32		&$-$0.26		&d\\
	&(0.002)	&(0.03)		&(0.2)	&(0.2)	&(0.05)	&	(0.003)&	(0.03)&		(0.05)&		(0.05)&\\
I	&12 30 48.709	&12 23 31.74	&0.49	&0.49	&11.03		&12 30 48.703	&12 23 31.82	&$-$0.08		&+0.08		&c,d\\
	&(0.004)	&(0.05)		&(0)	&(0)	&(0.08)	&	(0.001)&	(0.01)&		(0.05)&		(0.05)&\\
A	&12 30 48.621	&12 23 32.29	&0.98	&0.91	&12.43		&12 30 48.616	&12 23 32.41&	$-$0.07		&+0.11&\\
	&(0.001)	&(0.02)		&(0.04)	&(0.04)	&(0.03)	&	(0.001)&	(0.01)&		(0.02)&		(0.02)&\\
B	&12 30 48.505	&12 23 32.87	&1.7	&1.3	&14.2		&12 30 48.495	&12 23 32.99	&$-$0.14		&+0.13&\\
	&(0.006)	&(0.07)		&(0.2)	& (0.2) &(0.1)	&	(0.001)&	(0.01)&		(0.08)&		(0.08)&\\
C$-$1	&12 30 48.302	&12 23 34.56	&0.49	&0.49	&17.6		&12 30 48.296	&12 23 34.64&	$-$0.09		&0.09		&c\\
	&(0.005)	&(0.07)		&(0) 	&(0)	&(0.1)	&	(0.001)&	(0.02)&		(0.07)&		(0.07)&\\
C$-$2	&12 30 48.211	&12 23 34.94	&0.49	&0.49	&18.98		&12 30 48.256	&12 23 34.68	&+0.66		&$-$0.25&\\
	&(0.004)	&(0.06)		&(0.)	&(0)	&(0.09)	&	(0.001)&	(0.01)&		(0.07)&		(0.07)&
\tablenotetext{a}{Both the {\it HST} and {\it Chandra} data were registered to the {\it VLA} data assuming a common position for the nucleus of M87 
in the radio, optical and X-rays.  The position given for the nucleus is therefore based on the absolute {\it VLA} astrometry and all other positions
are relative to it.}
\tablenotetext{b}{Optical position for knot HST$-$1 was based on fitting the unconvolved HST image with a Gaussian of 
FWHM 0.174 arcsec, which was taken as representative of the resolution of the HST data with this pixellation.
The fit for the convolved HST image failed because of confusion with the nucleus.}
\tablenotetext{c}{Fits with a variable component size failed; a fixed FWHM of 1 ACIS pixel ($0\farcs492$) was assumed.}	
\tablenotetext{d}{Component sizes based on a free component size but component position is based on a fixed 0\farcs492 FWHM component to
      minimize error.}

\enddata
	
\end{deluxetable*}
\clearpage	

\begin{deluxetable*}{cclccccc} 
\tablecolumns{8}
\tablewidth{0pt}
\tablecaption{X-ray Spectral Fits for Nucleus and Jet Knots}

\tablehead{
\colhead{Component}&\colhead{Region}&\colhead{Distance}&\colhead{$\alpha_x$}&\colhead{$N_H^a$} & \colhead{$K^{b}$} & \colhead{S (1 keV), $\mu$Jy}&\colhead{$\chi^2_\nu$}
}

\startdata

Nucleus& 	$0\farcs6^c$ circle & 0 & 1.23 $\pm$ 0.11 & $3.5 \pm 1.5$ 	 & 17.68   & 107 & 0.933/68 \\

HST-1 &		$0\farcs6^c$ circle& $1\farcs25^d$ & $1.32 \pm 0.08$ & $2.4 \pm 0.9$	 & 20.76   & 129 & 0.968/77 \\

D-E &	  	$0\farcs6^c$ circle & $2\farcs91$ & 1.43 $\pm$ 0.09 & 2.4 (frozen)     	 & 8.31    & 51.5& 0.745/34 \\

E&	  	$2\farcs4 \times 1\farcs6 $ box &  $5\farcs80$& 1.48 $\pm$ 0.12 & $2.1 \pm 1.1$	  & 5.04    & 32.2& 0.618/64 \\

F	&	  	$2\farcs4 \times 1\farcs6 $ box & $  7\farcs75$	& 1.64 $\pm$ 0.15 & $1.7 \pm 1.3$    	 & 3.70    & 20.1& 0.760/42 \\

A&	  	$2\farcs3 \times 2\farcs0 $ box	&$12\farcs6$	& 1.61 $\pm$ 0.07 & $0.8 \pm 0.5$    	 & 27.7     & 156&  1.198/89 \\

B&		$3\farcs3 \times 1\farcs8 $ box	&$15\farcs35$  	& $1.59 \pm 0.12$ & $2.8 \pm 1.2$    	 &  5.70    & 30.3 & 0.678/58  \\

C&		$7\farcs8 \times 1\farcs5 $ box	& $20\farcs9^e$  	& 1.33 $\pm$ 0.06 & 2.4 (frozen)    	 &  3.58    & 20.6 &  1.576/49  \\
\enddata
\tablenotetext{a}{$10^{20} {\rm ~cm^{-2}}$}
\tablenotetext{b}{$10^{-5} {\rm ~Photons ~sec^{-1} ~keV^{-1}}$}
\tablenotetext{c}{Radius of extraction circle}
\tablenotetext{d}{Offset by $0.3''$ away from component centroid to avoid contamination from wings of nuclear PSF.}
\tablenotetext{e}{Offset by $2''$ away from nucleus in order to include any flux further out in jet.}
\end{deluxetable*}

\end{document}